\documentclass[reprint, amsmath,amssymb,aps,prfluids,showkeys,floatfix]{revtex4-2}
\usepackage{graphicx}
\begin{document}

\title{Hydrodynamics of tandem flapping pectoral fins with varying stroke phase offsets}

\author{Kaushik Sampath}\email[]{kaushik.sampath@nrl.navy.mil}\affiliation{Acoustics Division, Code 7165, U.S. Naval Research Laboratory, Washington, DC}
\author{Jason D Geder}\author{Ravi Ramamurti}\affiliation{Laboratories for Computational Physics and Fluid Dynamics, Code 6041, U.S. Naval Research Laboratory, Washington, DC}
\author{Marius D Pruessner}\author{Raymond Koehler}\affiliation{Center for Bio/Molecular Science \& Engineering, Code 6930, U.S. Naval Research Laboratory, Washington, DC}
\date{\today}

\begin{abstract}
We show how phasing between tandem bio-inspired fins flapping at high stroke amplitudes modulates rear fin thrust production and wake characteristics. Load cell thrust measurements show that the rear fin generates 25\% more thrust than the front fin when it lags the latter by a quarter cycle, and performs 8\% worse when it leads the front fin by the same amount. The flow interactions between the fins responsible for these observations are analyzed using 2-D particle image velocimetry (PIV) measurements and 3-D computational fluid dynamics (CFD) simulations. Distributions of velocity elucidate variations in the effective flow induced on the rear fin for different phase offsets. Vortex structure interactions and particle rakes reveal the contributions of the leading- and trailing-edge vortices shed by the front fin in modulating the suction at the rear fin leading edge. Furthermore, the wake structure far downstream of the fins changes in its coherence, axial and radial extents for the different phase offsets. These findings are relevant for the design and performance optimization of various unmanned underwater vehicles that utilize such tandem systems.
\end{abstract}


\maketitle

\section{Introduction}
Research to identify the principles of fish locomotion and characterize the propulsive performance for a variety of artificial, bio-inspired underwater propulsion systems has steadily grown over the past few decades. To address the need for more effective and efficient maneuvering in marine environments, propulsion and control systems inspired by fish and other aquatic organisms are starting to provide viable alternatives to traditional vehicle thrusters and control surfaces in a range of underwater regimes. Of the biological organisms studied and propulsion systems developed, the predominant focus has been on flapping fins or foils, in various configurations, to achieve thrust.

Low-speed maneuvering and station keeping capabilities for small unmanned underwater vehicles (UUVs) is of great interest to the maritime community for operations in confined, dynamic-flow, shallow-water environments. To address these needs, several researchers have taken inspiration from the pectoral fins employed by various fish species to develop flapping mechanisms ~\cite{hobsonPilotFishMaximizingAgility1999,lichtDesignProjectedPerformance2004,zhouDesignControlBiomimetic2008,sitorusDesignImplementationPaired2009,katoBiomechanismsSwimmingFlying2008,palmisanoRoboticPectoralFin2012,mooredInvestigatingThrustProduction2008,tangorraEffectFinRay2010,gederDevelopmentUnmannedHybrid2017,RazorAUVAC,NaronauticalRobot}. In many of these fins, large amplitude rotational stroke motions are used to achieve thrust. Licht et al.~\cite{lichtDesignProjectedPerformance2004} demonstrated that at zero forward speed, mean thrust for a pitching and flapping (vertical stroke plane) is primarily a function of maximum flap angle velocity and performed tests between stroke amplitudes, $\Phi$ = 20-60$^{\circ}$. Sitorius ~\cite{sitorusDesignImplementationPaired2009} demonstrated forward speed capabilities for a rowing fin motion with $\Phi$ = 30-60$^{\circ}$. Similar amplitude flapping motions have been employed in other studies as well ~\cite{katoBiomechanismsSwimmingFlying2008,palmisanoRoboticPectoralFin2012,bandyopadhyayRelationshipRollPitch2012}. While the development of these flapping pectoral fins and their integration onto unmanned systems has demonstrated promising results for low-speed maneuvering, it is important to understand the flow effects between fins actuated in closed proximity as they are on small underwater vehicles.

There is a large amount of literature on the study of the swimming mechanisms of various fish species, including some on the effects of multi-body interactions between individual fins. Biologists have observed the coordinated body and fin motions exhibited by these organisms, identifying phase relationships~\cite{arreolaMechanicsPropulsionMultiple1996} and analyzing wake interactions~\cite{hoveBoxfishesTeleosteiOstraciidae2001,lauderLearningFishKinematics2006,flammangVolumetricImagingFish2011,druckerLocomotorFunctionDorsal2001} between moving surfaces. Studies to develop robotic fins inspired by nature have included analysis of the effects of fin shapes, stroke parameters, configurations, materials and surface curvature control techniques on thrust and propulsive efficiency~\cite{lichtDesignProjectedPerformance2004,zhouDesignControlBiomimetic2008,sitorusDesignImplementationPaired2009,katoBiomechanismsSwimmingFlying2008,palmisanoRoboticPectoralFin2012,mooredInvestigatingThrustProduction2008,tangorraEffectFinRay2010,barrettDragReductionFishlike1999,espositoRoboticFishCaudal2012}. 
The configuration of tandem, identical geometry fins flapping perpendicular to the direction of flow is similar to lift-based pectoral fin motions of some fish species~\cite{fishDiversityMechanicsPerformance2006}. This configuration is also found on a number of vehicles designs due to the control authority it affords in low-speed maneuvers~\cite{lichtDesignProjectedPerformance2004,gederDevelopmentUnmannedHybrid2017,RazorAUVAC,NaronauticalRobot}. 

Although much of this research on fin development has focused on the design and testing of individual fins, more recent studies have tried to understand multi-fin interactions between artificial propulsion and control surfaces similar to those seen in nature. Multiple studies have examined flow interactions between in-line pitching and heaving foils in 2D numerical simulations demonstrating the effects of the heave phase offset~\cite{akhtarHydrodynamicsBiologicallyInspired2007,rivalVortexInteractionTandem2011,broeringNumericalInvestigationEnergy2012} and foil spacing~\cite{broeringNumericalInvestigationEnergy2012} on thrust and propulsive efficiency. Further, Rival et al.~\cite{rivalVortexInteractionTandem2011} carried out particle image velocimetry (PIV) measurements for characterizing the flow in 2D pitching and heaving foils demonstrating agreement with the 2D numerical results. For a larger selection of phase offsets and foil spacings, Boschitsch et al.~\cite{boschitschPropulsivePerformanceUnsteady2014} used PIV to examine the performance of 2D pitching foils. Kurt and Moored~\cite{kurtFlowInteractionsTwo2018} experimentally compared 2D and 3D pitching foils including PIV analysis demonstrating that the optimal pitch phase offset, thrust gains, and efficiency gains for the two cases were different. However, they do not consider a heave or flapping motion. Recently, for a set of dorsal, anal, and caudal fins in a robotic bluegill sunfish, Mignano et al.~\cite{mignanoPassingWakeUsing2019} experimentally showed increased thrust and reduced lateral forces with proper relative fin position and stroke phasing with flapping amplitudes between $\Phi$ = 17-25$^{\circ}$ in a 3D setting. The above-mentioned studies consider small or non-existent heave amplitudes, $h$, where $h/c \leq$0.5, with $c$ being the chord length, and hence may not be directly applicable to larger-amplitude flapping pectoral fin vehicles that are of interest in this study~\cite{lichtDesignProjectedPerformance2004,sitorusDesignImplementationPaired2009,katoBiomechanismsSwimmingFlying2008,palmisanoRoboticPectoralFin2012,gederDevelopmentUnmannedHybrid2017,RazorAUVAC}. However, due to the lack of studies that do consider larger $h/c$ configurations, we will still discuss the flow physics reported in the above-mentioned studies in the context of the present study subsequently. 

Warkentin et al.~\cite{warkentinExperimentalAerodynamicStudy2007} performed a systematic study of thrusts, lifts and propulsive efficiencies for tandem flapping membrane wings with a substantially larger heave amplitude ($h/c$=1.26). They find that under certain phase angles, the tandem arrangement outperforms the single wing in terms of its efficiency by almost twice as much. Our group has also performed multi-fin studies with larger heave amplitudes in the past. In 3D computational studies, Ramamurti et al. investigated the effects of fin position and stroke phase offset between tandem pectoral fins ~\cite{ramamurtiComputationalFluidDynamics2018} and between pectoral and caudal fins~\cite{ramamurtiPropulsionCharacteristicsFlapping2019} where $\Phi$ = 20-55$^{\circ}$. Geder et al.~\cite{gederUnderwaterThrustPerformance2017} experimentally investigated the effects of tandem pectoral fin parameters on stroke averaged thrust for that same flap amplitude range. 

Motivated by the findings from Warkentin et al.~\cite{warkentinExperimentalAerodynamicStudy2007} and our prior efforts to experimentally characterize stroke-averaged forces for tandem fin systems~\cite{ramamurtiComputationalFluidDynamics2018,gederUnderwaterThrustPerformance2017}, we use experimental thrust measurements, 2-D PIV and 3-D CFD simulations to analyze the flow structures generated by two pectoral fins flapping in close proximity at varying fin stroke phase offsets.  The objective is to understand the flow mechanisms that control fin performance, based on which design of multi-fin systems with large heave amplitudes can be optimized and their implications on wake characterstics.

\section{Methods}
\subsection{Experimental Setup}
\begin{figure}[ht]
\begin{center}\includegraphics[width=\linewidth]{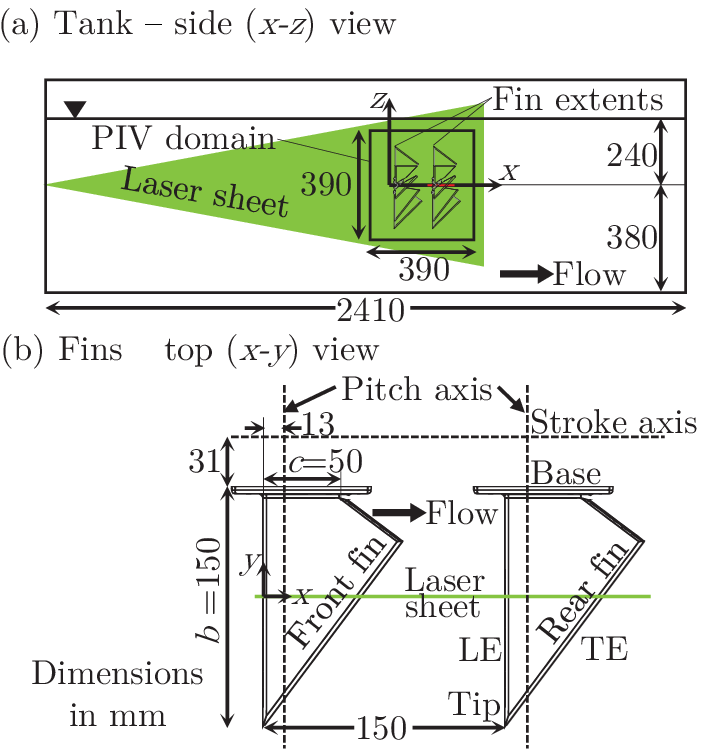}\end{center}
\caption{(a) Schematic of experimental setup showing side/axial-vertical (x-z) view of the rectangular glass tank. Origin in this view is defined at the intersection of the leading edge of the front fin at a zero stroke angle (vertical center of the PIV domain). (b) Top view of the setup highlighting the geometry and arrangement of the front and rear fins. Leading (LE) and trailing edges (TE), as well as the tip of the trail fin are labeled for clarity. Origin is defined as the intersection of the lead fin LE with the mid-span (laser sheet). Flow arrows indicate direction of front-fin induced flow.}
\label{fig_schematic}
\end{figure}

\begin{figure}[ht]
\begin{center}\includegraphics[width=\linewidth]{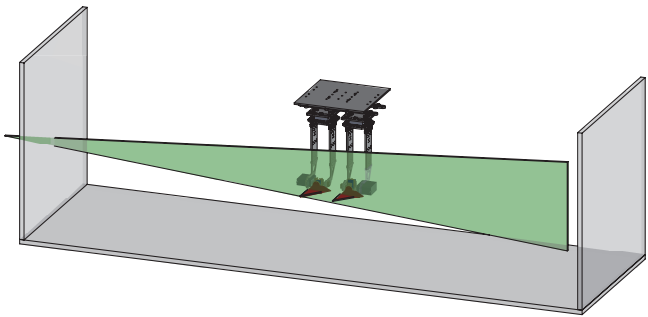}\end{center}
\caption{3D CAD view of the experimental setup including the tank, laser sheet and fin orientations.}
\label{fig_3DCAD}
\end{figure}

Experiments have been performed in a 2410 $\times$ 760 $\times$ 760 mm (length $\times$ width $\times$ height) transparent glass tank with 15.8 mm thick walls (Figure~\ref{fig_schematic}a). 
The dimensions along the length, width and height of the tank are denoted $x$ (axial), $y$ (lateral) and $z$ (vertical) respectively. The pectoral fin located on the left, hereby referred to as the `front fin' generates flow impinging on the `rear fin', located downstream (to the right in Figure~\ref{fig_schematic}a) along the length of the tank. The origin is defined at the intersection of the mid-span and the front fin leading edge. The axial distance between the leading edges of the two fins, $d$ is set to 150 mm ($d$ = 3$c$). A 3D sketch of the experimental setup, including the tank, laser sheet and fins is shown in Figure~\ref{fig_3DCAD}. 
\subsection{Fin design, control and kinematics}
Based on our prior experience~\cite{ramamurtiComputationalFluidDynamics2018,ramamurtiPropulsionCharacteristicsFlapping2019,gederDevelopmentUnmannedHybrid2017} and literature~\cite{walkerKinematicsDynamicsEnergetics2002,lauderLearningFishKinematics2006} we have opted for a bio-inspired trapezoidal fin with an aspect ratio of 3:1 (Figure~\ref{fig_schematic}b). This geometry produced greater thrust in a tandem two-fin configuration than an identical size and aspect ratio rectangular fin~\cite{ramamurtiPropulsionCharacteristicsFlapping2019}. 
 
The fins are mounted on independent frames that connect them to servo motors, Hitec D840WP and Hitec D646WP for stroke and pitch angle control respectively. Potentiometers (TT Electronics P260, 10k$\Omega$) present on each fin are used to measure and verify the kinematics. Two separate 3-axis force cells (Interface 3A60A, 20N) are used for measuring the thrust forces on each fin with an uncertainty in measurement of 0.04 N.

The fin has a mean and base chord length, $c$ = 50 mm and a span, $b$ = 150 mm. The fin stroke angle is varied over a stroke amplitude, $\Phi$ = 57.5$^{\circ}$ about an axis that is located 31.25 mm from the base of the fins parallel to the X-axis. This results in a fin tip stroke radius, $R$ = $b$ + 31.25 = 181.25 mm. Hence, the total arc length traversed in half a stroke is $L$ = 2$\pi\Phi$R/360$^{\circ}$ = 363.8 mm. 

\begin{figure}[ht]
\begin{center}\includegraphics[width=\linewidth]{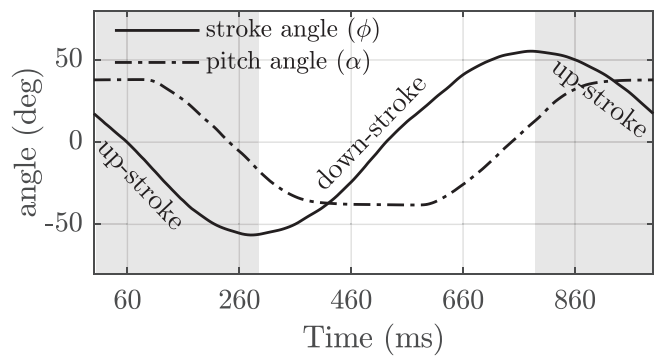}\end{center}
\caption{Cyclic variation of stroke ($\phi$) and pitch angle ($\alpha$). $\phi$ is positive when the fin is stroking downward in $z$ and $\alpha$ is positive when the fin leading edges are pitching upward in $z$. $x$-axis gridlines represent the ten time-steps for which PIV data is acquired per cycle.}
\label{fig_strokepitch}
\end{figure}

The pitch is varied from -40$^{\circ}$ to 40$^{\circ}$ about an axis parallel to the Y-axis, located inside the fin, at a distance, $c$/4 = 12.5 mm from the corresponding fin leading edge. The cyclic variation of stroke ($\phi$) and pitch angles ($\alpha$) is shown in Figure~\ref{fig_strokepitch}. 
The sign of the angles follows from the right-hand thumb rule applied to the respective rotation axis, i.e. $\phi$ is positive when the fin is stroking downward in $z$ and $\alpha$ is positive when the fin leading edges are pitching upward in $z$. The transition period during the $\phi$ sign reversal is around 300 ms long.  An offset angle of the rear fin stroke w.r.to that of the front fin is denoted by $\phi_{\text{offset}}$, based on which the stroke phase offset is defined as $\delta$ = ($\phi_{\text{offset}}$/$\Phi$).180$^{\circ}$. 

The present tests are performed with no external flow, i.e., $U_{\infty}$=0 m/s past the fins. Consequently, the Strouhal number, defined as, $St$ = $fL/U_{\infty}$ $\rightarrow\infty$, where $f$ = 1 Hz, denotes the flapping frequency. A zero free-stream flow condition is an important operating point for hover-capable underwater systems as they operate in quasi-steady flow environments with slow acceleration from a state of rest~\cite{lichtDesignProjectedPerformance2004,parkVorticalStructuresFlexible2016,shindeFlexibilityFlappingFoil2014}. Other relevant situations include station holding in shallow waters and hovering while offseting buoyency, i.e. generating a downforce to hold position.
\subsection{PIV measurements}
A 532 nm Nd:YAG dual-pulse laser with an energy of 70 mJ/pulse (EverGreen 70) is used to generate a thin light sheet in the $x$-$z$ plane that intersects the fins at their mid-span ($y$ = 0), refer Figure~\ref{fig_schematic}a. A dual exposure 2048$\times$2048 pixel CCD camera (TSI PowerView 630059) in combination with a 50 mm lens is used to capture two successive 390$\times$390 mm images of the light-sheet illuminated region centered about the tandem fins. The time interval between successive exposures is set to 2 ms. 

The fin mounting hardware (Figure~\ref{fig_3DCAD}) is present in the camera’s field of view. Although painted black and present substantially outside the focal plane, reflections of the light sheet from the hardware infiltrate the particle images acquired, especially the second (typically longer) PIV exposure. To address this issue, the tank is seeded with 38-45 $\mu$m diameter fluorescent particles (Cospheric UVPMS-BO-1.00) that are neutrally buoyant in water. A 560 nm long-pass optical filter is used to allow light only from the fluorescence of these particles to enter the sensor. This significantly improves the signal to noise ratio of the particle images acquired.

To facilitate ensemble averaging of the measured flow field for each stroke angle, phase-locked PIV measurements are performed. The term ‘phase-locked’ PIV is generally used in turbo-machinery to describe repeated PIV measurements performed at a certain blade phase while the blades are rotating. To clarify, in the present context, this actually refers to $\phi$– locked PIV measurements and has nothing to do with the stroke phase offset, $\delta$. Between 1300 – 2000 PIV image pairs are acquired for ten equally spaced values of $\phi_{front}$ corresponding to cycle time-steps of t = 60, 160...960 ms for all three $\delta$ values. For convenience, these time-steps are displayed as $x$-axis markers in Figure~\ref{fig_strokepitch}. 

\begin{figure}[ht]
\begin{center}\includegraphics[width=\linewidth]{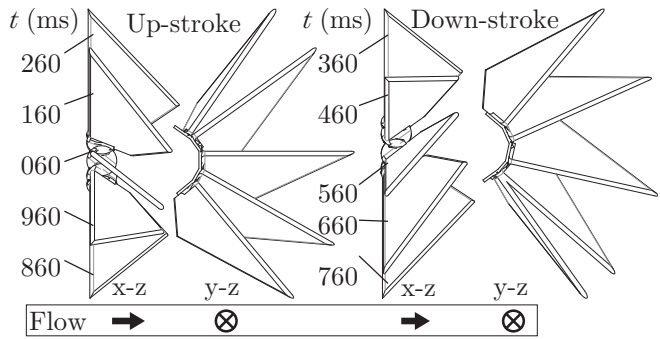}\end{center}
\caption{Views in the $x$-$z$ and $y$-$z$ planes of the front fin for ten time-steps in a single cycle when PIV data was recorded. For convenience, times corresponding to up- and down-stroke are shown in the left and right columns respectively. Flow arrows indicate direction of front-fin induced flow.}
\label{fig_finviews}
\end{figure}

The axial and vertical flow velocity components are denoted $u$ and $w$ respectively. The ensemble stroke-cycle averaged distributions of  $u$ and $w$ are denoted as $U$ and $W$.The axial and vertical distances are normalized by $c$. The velocities are normalized by the average tip speed, $V_{tip}$ = 2$Lf$  = 0.72 m/s. The ensemble-averaged spanwise vorticity is defined as, $\Omega_y$ = $\partial W/\partial x -\partial U/\partial z$. The front fin orientations for all ten values of $\phi_{front}$ in the $x$-$z$ and $y$-$z$ planes are shown in Figure~\ref{fig_finviews}.
Orientations corresponding to the up- and down-strokes are shown in the left and right columns respectively. Further details of PIV data processing and uncertainty estimates are provided in Appendix~\ref{AppA}.

\subsection{CFD simulations}

The governing equations employed are the incompressible Navier-Stokes equations in Arbitrary Lagrangian-Eulerian (ALE) formulation which are written as:
\begin{equation}
\frac{d \bf v}{dt} + \bf v_a.\nabla \bf v + \nabla p = \nu \nabla^2\bf v
\end{equation}
\begin{equation}
\nabla.\bf v = 0
\end{equation}

Here, $p$ denotes the pressure, $\nu$ is the kinematic viscosity, $\bf v_a$ = $\bf v$ – $\bf w$, the advective velocity vector, where $\bf v$ is the flow velocity and $\bf w$ is the mesh velocity and the material derivative is with respect to the mesh velocity $\bf w$ (not to be confused with the vertical flow velocity $w$). Both the pressure $p$ and the viscous stress tensor are normalized by the (constant) density $\rho$ and discretized in time using an implicit time stepping procedure. Thus, the equations are Eulerian for zero mesh velocity and Lagrangian if the mesh velocity is the same as the flow velocity. The time-accurate flow solver is discretized in space using a Galerkin procedure with linear tetrahedral elements. General details of the flow solver have already been discussed extensively elsewhere~\cite{ramamurtiEvaluationIncompressibleFlow1992,ramamurtiComputation3DUnsteady1999} in connection with successfully validated solutions for several 2-D and 3-D, laminar and turbulent, steady and unsteady flows.

Specific to this study, approximately 2 million points and 12 million tetrahedral unstructured elements are used to solve for every case. Snapshots showing the CFD grid resolution for the $\delta$ = -90$^o$ case around the two fins as well zoomed into the front fin LE are shown in Figure~\ref{fig_CFDresolution}. The Reynolds number based on $V_{tip}$ is 36,500, identical to that in the PIV measurements, and based on its magnitude is deemed to be in the laminar regime. The simulation time-step is approximately 10$^{-4}$ s and variable. The 3-D unsteady CFD simulations are carried out for the same fin kinematics as that of the PIV and thrust experiments, with a temporal resolution of 100 angles per stroke cycle for multiple cycles. Axial thrust profiles are computed from the unsteady surface pressure distribution on the fins and not the wakes. They would however, include the pressure changes due to separation and reattachment (if any).


\section{Results and Discussion}
\subsection{\label{thrustprofiles}Thrust and Power}

For a given axial distance between flapping fins, stroke amplitude and frequency, there appears to be an advantage in terms of the thrust output of the rear fin when  $\delta$ is varied~\cite{ramamurtiComputationalFluidDynamics2018,ramamurtiPropulsionCharacteristicsFlapping2019,gederDevelopmentUnmannedHybrid2017}. Hence, we look at thrust and power measurements at several $\delta$ values to probe the same. Measured rear fin thrust profiles averaged over multiple flapping cycles, denoted as $\langle T\rangle$, for a range of $\delta$ values are shown in Figure~\ref{fig_thrust_delta} along with the corresponding results from CFD. They show that $\delta$ = -90$^o$ and +90$^o$ correspond to the maximum and minimum rear fin thrust respectively, while thrust is identical to that of the front fin (not shown) at 0$^o$. These results are in agreement with our prior findings for different fin kinematics~\cite{ramamurtiComputationalFluidDynamics2018,ramamurtiPropulsionCharacteristicsFlapping2019,gederDevelopmentUnmannedHybrid2017}. 

It is also important to understand the relationship between $\delta$ and power consumption to determine if higher or lower propulsive efficiencies might be achieved at different $\delta$. The total power input, $P$, can be expressed as $P$ = $P_m$+ $P_h$, where $P_m$ and $P_h$ are the mechanical and hydrodynamic powers respectively. The rear fin power across a range of $\delta$ values (Figure~\ref{fig_power}) demonstrates a negligible (2\%) variation in $P$ with $\delta$. Measurements of $P$ with the fins running inside an empty tank ($P\sim P_m$ = 6.0 W) showed a reduction of 16\% compared to when the tank is filled with water ($P$ = $P_m$+$P_h\sim$ 7.2 W) indicating the dominance of $P_m$ in $P$. 

Moreover, CFD computed $P_h$ values lie in the 0.82-0.97 W range for the rear fin, in reasonable agreement with the estimates from the experiments ($\sim$1.2 W). This indicates that even 25\% variations in $P_h$ amount to a neglibile 4\% change in $P$. Therefore, for the present study, it is assumed the desired operating $\delta$ is not influenced by power considerations. While changes in $P_h$ are negligible compared with $P$, it is worth noting that a change in $h/c$ or other fin parameters induces a substantial change in $P_m$ and thus $P$, affecting vehicle performance.

CFD analysis was also performed for $U_\infty$ = 0.5 m/s ($Re$ = $cU_\infty/\nu$ = 25,000 and $St$ = 0.73), and the resulting $\langle T\rangle$ profiles for the front and rear fins at various $\delta$ values are shown in Figure~\ref{fig_Uinf}. Notably the thrust magnitudes are significantly lower than those for the zero flow case in agreement to prior findings~\cite{ramamurtiComputationalFluidDynamics2018,ramamurtiPropulsionCharacteristicsFlapping2019}. The incoming flow reduces the incident angle of attack, thereby decreasing the thrust generated by the fins. However, it is still positive implying that the vehicle drag can still be overcome to go forward while experiencing an in-flow of 0.5 m/s. 

The relative trends (w.r.to $\delta$) however, remain similar to those for the zero flow case. There is no apparent effect of $\delta$ on the front fin thrust, while the rear fin is able to leverage from the front fin wake structure and generate higher thrust at $\delta$=0$^o$, now advanced by 90$^o$ compared to the static case (Figure~\ref{fig_thrust_delta}). Consequently, the minimum rear fin thrust occurs at -90$^o$. This shift in the optimum $\delta$ is due to the incoming flow accelerating the arrival of the front fin wake on the rear fin, but the interaction remains of a similar nature. We defer discussion on other interesting flow features that differentiate the $U_\infty$ = 0.5 m/s case from the zero in-flow case to future studies and restrict present analysis to characterizing the effect of $\delta$ over the $U_\infty$=0 cases.

Therefore, we select the $\delta$ = -90$^{\circ}$, 0$^{\circ}$ and +90$^{\circ}$ cases at zero in-flow for subequent time-series analysis, as well as for performing PIV measurements as they cover the entire range of rear fin thrust leveraged by $\delta$ (Figure~\ref{fig_thrust_delta}). 

\begin{figure}[ht]\begin{center}\includegraphics[width=\linewidth]{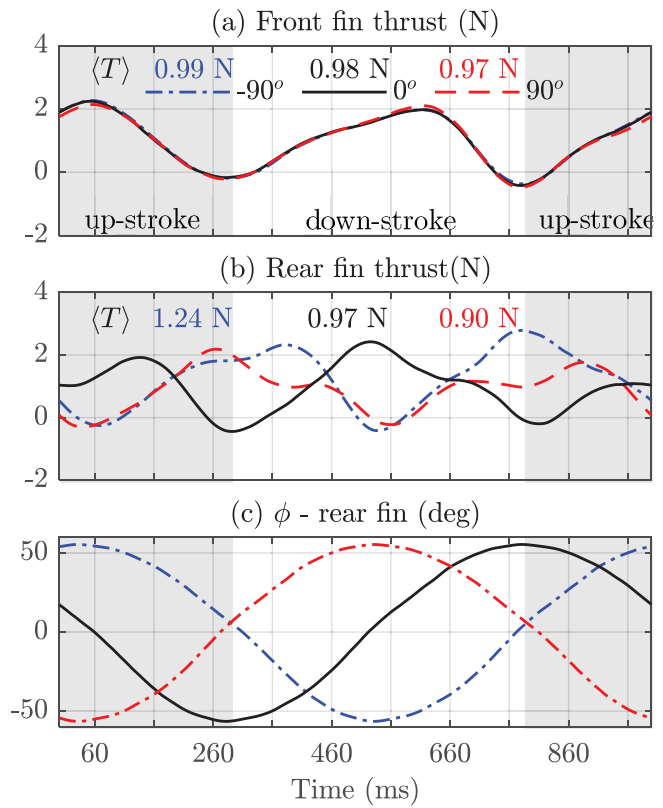}\end{center}
\caption{Measured thrust profiles at the (a) front and (b) rear fins at stroke phase offsets $\delta$ = -90$^{\circ}$, 0$^{\circ}$ and +90$^{\circ}$.  Average thrust value, $\langle T\rangle$ is provided at the top for each $\delta$ case. (c) Rear fin $\phi$ for stroke phase offsets $\delta$ = -90$^{\circ}$, 0$^{\circ}$ and +90$^{\circ}$. Note that the time axis is w.r.to the front fin, and the staggering of profiles in (b) corresponds to the staggering of $\phi$ in (c). A direct comparison of the rear fin thrust values for different upstream conditions is provided in Figure~\ref{fig_rearfinthrust}.}
\label{fig_thrustresults}
\end{figure}

The axial thrust obtained from the load cells at the front and rear fin is averaged over all the cycles acquired. A first-order lowpass filter is applied to remove high frequency fluctuations associated with the load cell element natural frequency (7 Hz) and sensor noise. The resulting profiles are shown in Figure~\ref{fig_thrustresults}. 
$\langle T\rangle$ is provided at the top for each $\delta$ case. 

\begin{figure}[ht]\begin{center}\includegraphics[width=\linewidth]{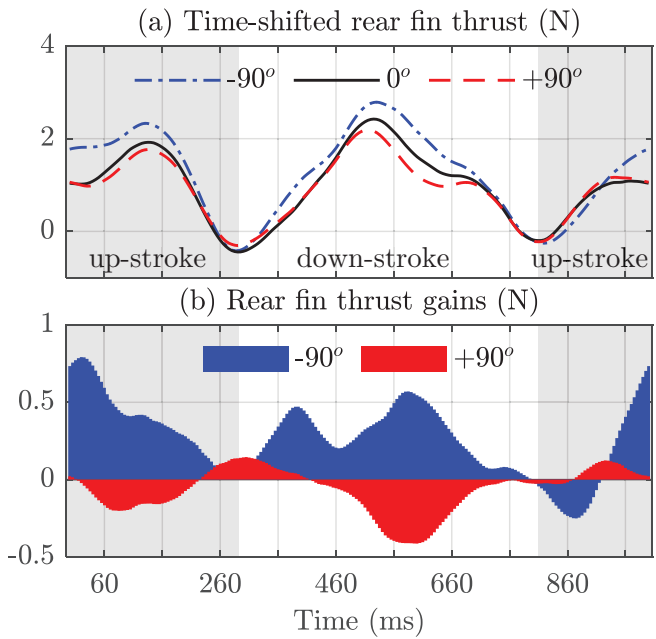}\end{center}
\caption{(a) Measured rear fin thrust profiles time-shifted to match the front fin stroke cycle for $\delta$ = -90$^{\circ}$, 0$^{\circ}$ and +90$^{\circ}$. (b) Measured rear fin thrust gains for $\delta$ = -90$^{\circ}$ and +90$^{\circ}$ w.r.to the $\delta$ = 0$^{\circ}$ case.}
\label{fig_rearfinthrust}\end{figure}

The front fin profiles (Figure~\ref{fig_thrustresults}a) for all $\delta$ values lie on top of each other, indicating that there is no time effect (either) of the rear fin phasing on the axial thrust produced at the front fin. Front fin thrust increases with stroke progression, peaks around mid-stroke and reaches a minimum value at reversal. The thrust profiles from the rear fins (Figure~\ref{fig_thrustresults}b) are staggered, due to the phasing. In order to make more sense and  unwrap phasing effects,  Figure~\ref{fig_rearfinthrust}a shows the same three profiles as those in Figure~\ref{fig_thrustresults}b now time-shifted so as to match the rear fin stroke cycles to that of the front fin.
Evidently, the $\delta$ = -90$^\circ$ case has the highest thrust, higher than the front fins by 25$\%$ followed by the in-phase $\delta$ = 0$^\circ$ case, while the $\delta$ = +90$^\circ$ performs $8-10\%$ worse than the front fins on average.

The average thrust over the cycle produced at the rear fin is the same as that of the front fin when they are in phase ($\delta$ = 0$^{\circ}$). Their peak values are also the same, at around 2 N. Hence, the in-phase case is subsequently treated as a `reference' to compare the performance of the $\delta$ = -90$^{\circ}$ and +90$^{\circ}$ cases and subtracted from them to further elucidate the points along the cycle where gains or losses in axial thrust are incurred by the rear fins in Figure~\ref{fig_rearfinthrust}b. The $\delta$ = -90$^\circ$ case gains almost throughout the strokes, with significant thrust enhancement during mid-strokes. The $\delta$ = +90$^\circ$ curve on the other hand, mostly performs worse than the reference case, resulting in a significant reduction in thrust when compared to the  $\delta$ = -90$^\circ$ curve, especially during mid-strokes. Although, there are times (near stroke inception) when the gains switch signs, i.e. $\delta$ = -90$^\circ$ performing worse than the $\delta$ = +90$^\circ$ case, the amplitudes are small and short-lived.

To put these rear fin thrust results in context with other multi-fin studies from the literature, Table~\ref{table001} contains their details sorted chronologically. It provides values of $Re$, based on in-flow ($cU_\infty/\nu$) as well as tip velocity ($cV_{tip}/\nu$), $St$, chord-normalized heave amplitude ($h/c$), front to rear fin separation ($d/c$), optimum phase offset ($\delta^*$) w.r.to the rear thrust for $d/c\sim3$ and the ratio between the rear and front fin thrust coefficients at that optimum phase offset, denoted $C_T^r/C_T^f$ at $\delta^*$.

\begin{table*}
\caption{\label{table001}Comparison of study parameters from various multi-fin studies.}
\begin{ruledtabular}\begin{tabular}{@{}lllllllll}
Study															&Study Type			&\multicolumn{2}{c}{$Re$ [$10^3$]}		&$St$			&$h/c$ 		&$d/c$ 		&-$\delta^{*}|_{d\sim3c}$	&$C_T^r/C_T^f|_{\delta^{*}}$\\
																&					&$cU_\infty/\nu$	&$cV_{tip}/\nu$		&$fL/U_\infty$	&		&		&						&\\
\hline
Akhtar et al.~\cite{akhtarHydrodynamicsBiologicallyInspired2007}					&2D CFD				&0.6				&0.3					&0.28		&0.56	&2		&48$^o$					&3\\
Warkentin et al.~\cite{warkentinExperimentalAerodynamicStudy2007}				&Force measurements	&29.7			&21.4-39.8			&0.36-0.67	&1.26	&1.4-3.7	&45-90$^o$				&1.3-1.7\\
Rival et al.~\cite{rivalRecoveryEnergyLeading2011,rivalVortexInteractionTandem2011}	&2D PIV, 2D CFD		&30.0			&3.0-4.8				&0.08, $\infty$	&0.50	&2		&90$^o$					&$\sim$1\\
Broering et al.~\cite{broeringNumericalInvestigationEnergy2012}					&2D CFD				&10.0			&6.0					&0.3			&0.50	&2		&0$^o$					&1.4\\
Boschitsch et al.~\cite{boschitschPropulsivePerformanceUnsteady2014}				&Force, 2D PIV			&4.7				&2.4					&0.25		&0		&1.25-5.25&150$^o$					&1.3\\
Ramananarivo et al.~\cite{ramananarivoFlowInteractionsLead2016}				&Force, flow visualization	&1.0-10.0			&0.6-2.0				&0.1-0.15		&0.1-0.8	&2-14	&0$^o$					&$\sim$1.3\\
Kurt and Moored~\cite{kurtFlowInteractionsTwo2018}							&Force, 2D PIV 			&7.5				&3.8					&0.25		&0		&1.25-2.25&30$^o$					&1.3-1.6\\
Lin et al.~\cite{linPhaseDifferenceEffect2019}								&2D CFD				&0.2				&0.1					&0.24		&0.40	&1.1-4.0	&60$^o$					&NA$|_{d\sim3c}$\\
{\bf Present study}													&Force, 2D PIV, 3D CFD	&0				&36.5				&$\infty$		&3.15	&3		&90$^o$					&1.25
\end{tabular}
\end{ruledtabular}\\
\end{table*}

Comparisons of the CFD results with the measured thrust profiles are detailed in Appendix~\ref{appendix_CFDValidation}.

Subsequent discussion utilizes results from the 2-D PIV measurements and 3-D CFD simulations for $U_\infty$=0, to look at the flow interactions that explain the effect of phasing on thrust profiles (Figure~\ref{fig_thrustresults}) cognizant of the flow mechanisms that have been linked to rear thrust performance in the past. Hereafter for brevity, we only focus on one of the strokes - upstroke (chosen arbitrarily) while examining the flow seen by the rear fins (time-shifted w.r.to the rear fins) in each of the three phase offsets.  

\subsection{Velocity distributions}
\label{Velocitydistributions}
\begin{figure*}[ht]
\begin{center}\includegraphics[width=\linewidth]{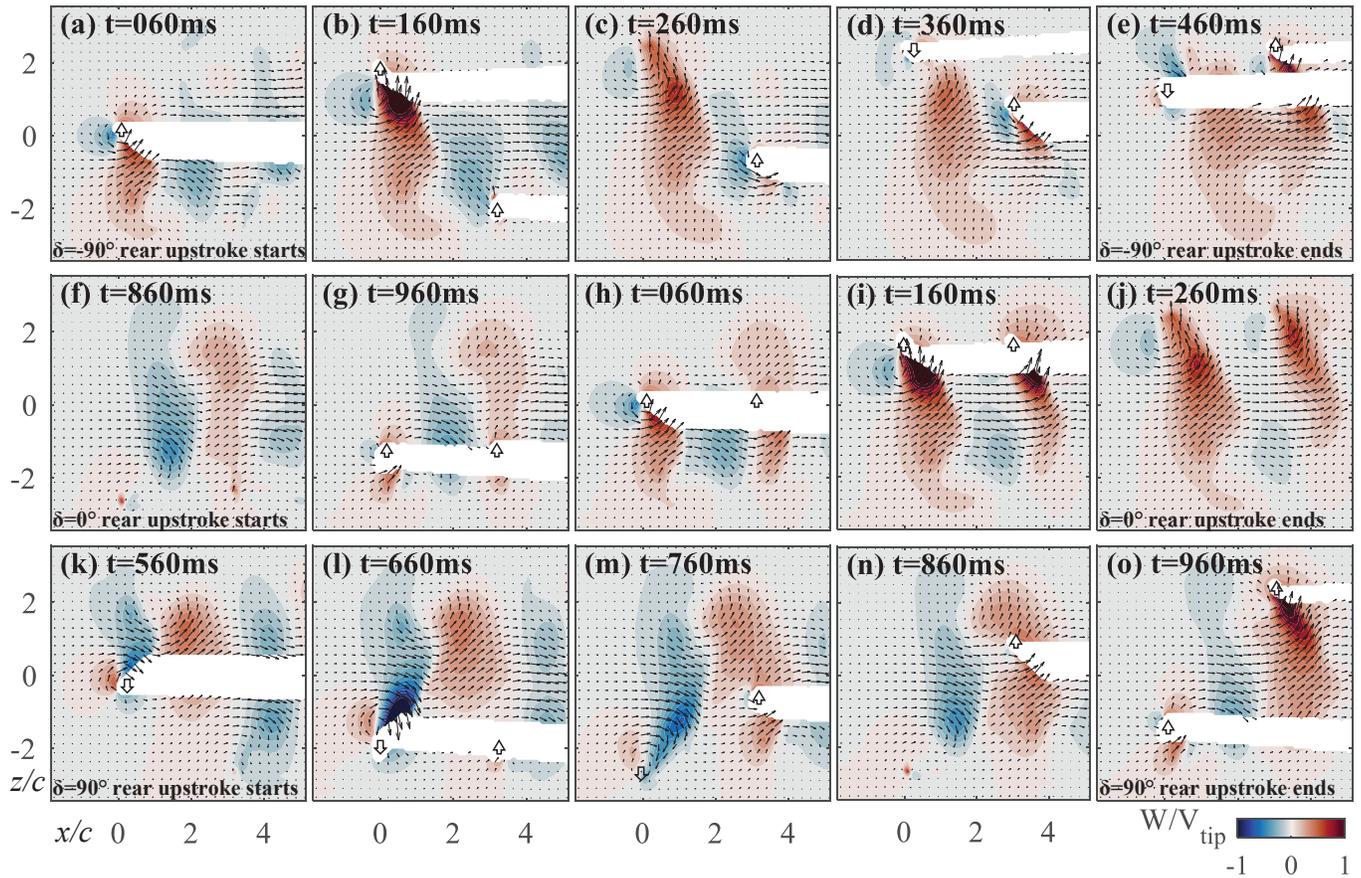}\end{center}
\caption{Contour distributions of $W/V_{tip}$ from PIV data overlaid with velocity vector field (every fifth vector is shown in each direction for clarity) for five time-steps spanning the rear fin upstroke at $\delta$ = -90$^{\circ}$ (a-e), 0$^{\circ}$ (f-j) and +90$^{\circ}$(k-o). Horizontal and vertical axes represent $x/c$ and $z/c$ respectively. Time-steps correspond to the front fin timing as shown in Figure~\ref{fig_thrustresults}. Arrows at the front and rear fins represent the direction of their vertical motion.}
\label{fig_Wcontour}
\end{figure*}

Following Boschitsch et al.~\cite{boschitschPropulsivePerformanceUnsteady2014}, we identify flow mechanisms from the velocity field and illustrate inflow assistance that the rear fin receives from the front for different $\delta$. Hence, contour distributions of  $W/V_{tip}$ from PIV data overlaid with the velocity vector field for five time-steps spanning the rear fin up-stroke at the three $\delta$ values is shown in Figure~\ref{fig_Wcontour}.
Every fifth vector is shown in each direction for clarity. The horizontal and vertical axes denote $x/c$ and $z/c$ respectively. The empty (vector-less) regions are either the in-plane cross-sections of the fins or the shadows exerted by the front fin. As per the definition introduced earlier, the front fin is located at $x/c$=0, while the rear fin is at $x/c$=3. The first, second and third rows represent $\delta$ = -90$^{\circ}$ (Figure~\ref{fig_Wcontour}a-e), 0$^{\circ}$ (Figure~\ref{fig_Wcontour}f-j) and +90$^{\circ}$(Figure~\ref{fig_Wcontour}k-o) respectively. The first column is just before the rear fin starts its upstroke in each of the cases (rear LE near $x/c$ = 3, $z/c$ = -2.5) and the last column is prior to the end of the upstroke (rear LE near $x/c$ = 3, $z/c$ = +2.5). Comparisons of the CFD results with PIV velocity profiles are detailed in Appendix~\ref{appendix_CFDValidation}.

The high-momentum wakes generated by the front fin are evident for all the snapshots and $\delta$ values. The wake reaches its peak magnitude near the front fin mid-stroke ($z/c$ = 0) and rapidly diffuses as it advects downstream by the next snapshot. This can be seen in Figure~\ref{fig_Wcontour}b-c, i-j and l-m for $\delta$ = -90$^{\circ}$, 0$^{\circ}$ and +90$^{\circ}$ respectively. By the time the rear fin reaches mid-stroke (Figure~\ref{fig_Wcontour}d, i, n), the front fin wake propagates to above, in-line and below the rear fin for the  $\delta$ = -90$^{\circ}$, 0$^{\circ}$ and +90$^{\circ}$ cases respectively. The wake is closer to the rear fin in the $\delta$ = -90$^{\circ}$ and +90$^{\circ}$ cases, but weaker in comparison to the in-phase case. Throughout its upstroke cycle, the rear fin experiences a substantially different $W$ from each of the three $\delta$ cases. In the $\delta$ = -90$^{\circ}$ case, the rear fin leading-edge (RLE) is associated with a downward (W$<$0) front fin flow, that increases its effective suction. For the in-phase case, W$\sim$0, suggesting no interaction between the two wakes, whereas W$>$0 for the $\delta$ = +90$^{\circ}$ case, acting against the incoming flow.

To quantify the above differences, Table~\ref{table002} contains the net angle of attack ($\alpha$), coefficients of lift ($C_L$), drag ($C_D$) and thrust ($C_T$), velocity induced ($V_{ind}/V_{tip}$) and $C_T (V_{ind}/V_{tip})^2$ (proportional to thrust) at $x/c=3$ when the rear-fin is at mid-upstroke for the three $\delta$ values.

\begin{table}[ht]
\caption{\label{table002}Front fin-induced flow at rear fin}
\footnotesize
\begin{ruledtabular}
\begin{tabular}{@{}llll}
$\bf \delta$				&\bf -90$^\circ$		&\bf 0$^\circ$	&\bf +90$^\circ$\\
\hline
$\bf \alpha$				&31$^\circ$			&35$^\circ$		&25$^\circ$	\\
$\bf C_L$ 				&0.79				&0.75			&0.84\\
$\bf C_D$ 				&0.51 			&0.55			&0.45\\
$\bf C_T$ 				&0.58				&0.58			&0.57\\
$\bf V_{ind}/V_{tip}$ 		&1.21				&1.15			&0.85\\
\hline
$\bf C_T (V_{ind}/V_{tip})^2$	&0.85				&0.78			&0.41\\
\end{tabular}
\end{ruledtabular}
\end{table}

The values of $C_D$, $C_L$ correspond to those for a flat plate taken from Ortiz et al.~\cite{ortizForcesMomentsFlat2015} for the corresponding $\alpha$. As evident from Table~\ref{table002}, $\alpha$ is roughly in the same range (25$^\circ$-35$^\circ$) for the three $\delta$ cases, resulting in similar $C_D$, $C_L$ and $C_T$. However, due to the opposing signs of $W$ induced by the front fin flow between the $\delta$ cases as described above, the values of $V_{ind}$ vary and result in lower thrust ($\sim C_T (V_{ind}/V_{tip})^2$) at the rear fin for the $\delta$ = +90$^\circ$ case compared to the other two. 

In other words, the former represents a foil with a lower incident flow, resulting in a decrease in thrust when compared to the $\delta$ = 0$^\circ$ case, whereas the $\delta$ = -90$^\circ$ case is a foil with a higher incident flow. This is indeed similar to the effective velocity enhancement and Knoller-Betz effect~\cite{knollerGesetzedesLuftwiderstandes1909,betzBeitragZurErklaerung1912} as described in detail by Boschitsch et al. \cite{boschitschPropulsivePerformanceUnsteady2014}.
\subsection{Vortex structure interactions}
Several  studies~\cite{akhtarHydrodynamicsBiologicallyInspired2007,rivalRecoveryEnergyLeading2011,rivalVortexInteractionTandem2011,boschitschPropulsivePerformanceUnsteady2014}, have focused on the impingement of the front fin-induced vortex structures on the rear fin as another tool to understand these multi-fin interactions due to phasing. For a pair of closely spaced line foils with a small $h/c$ = 0.56, 2-D CFD results from Akhtar et al.~\cite{akhtarHydrodynamicsBiologicallyInspired2007} show that the enhancement of thrust at the rear is associated with a negative (clockwise) front foil vortex impinging on the rear fin leading edge (RLE) as it moves up (and vice versa). Rival et al.~\cite{rivalRecoveryEnergyLeading2011,rivalVortexInteractionTandem2011} also concluded that a clockwise front leading edge vortex (FLEV) stays attached on the RLE during part of its upstroke (and vice versa), resulting in the observed thrust performance at the optimum phase offset. 

\begin{figure*}[ht]\begin{center}\includegraphics[width=\linewidth]{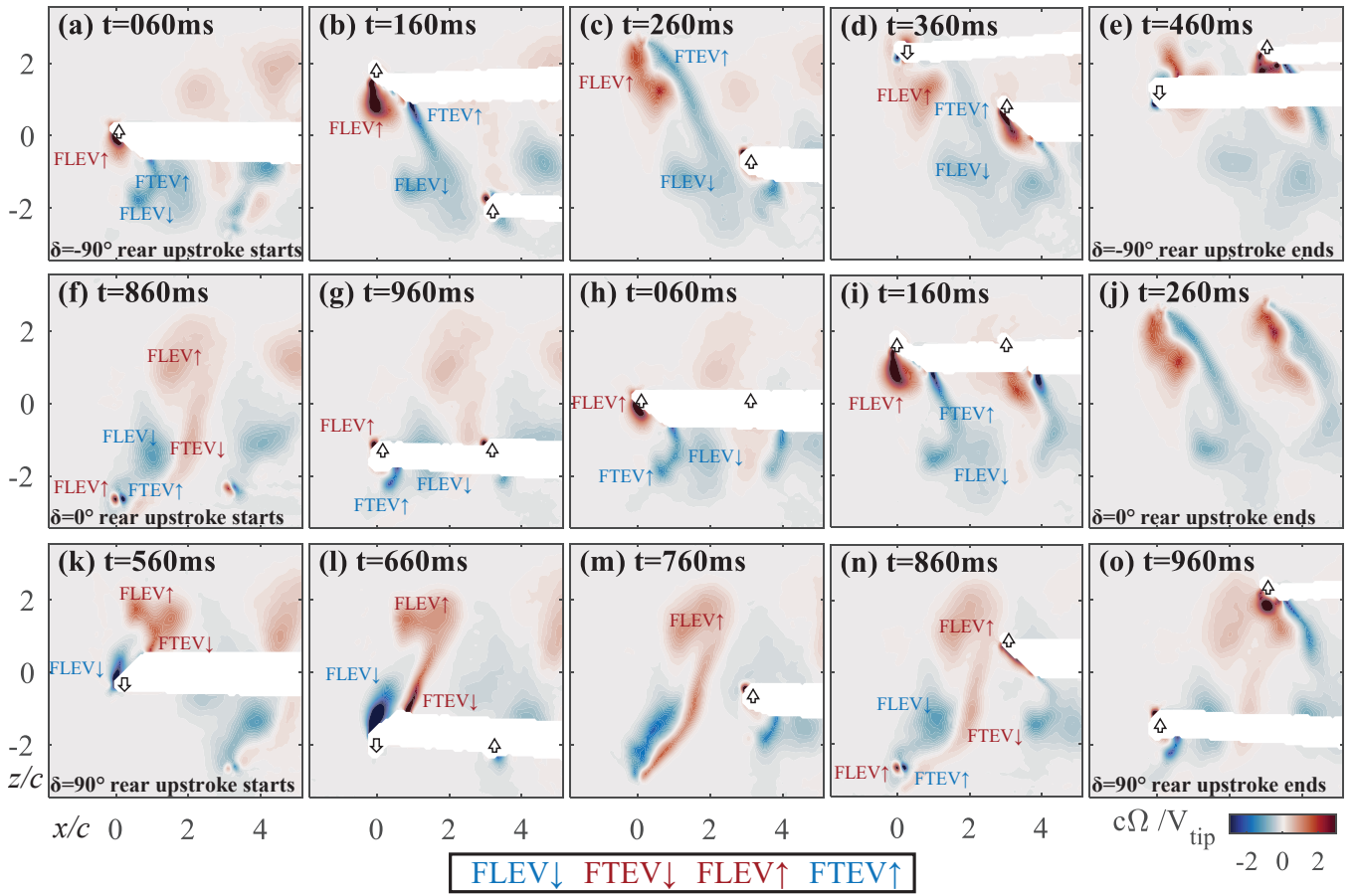}\end{center}
\caption{Contour distributions of $c\Omega/V_{tip}$ from PIV data for five time-steps spanning the rear fin upstroke at $\delta$ = -90$^{\circ}$ (a-e), 0$^{\circ}$ (f-j) and +90$^{\circ}$(k-o). Horizontal and vertical axes represent $x/c$ and $z/c$ respectively. Time-steps correspond to the front fin timing as shown in Figure~\ref{fig_thrustresults}. Arrows at the front and rear fins represent the direction of their vertical motion.}
\label{fig_Ocontour}\end{figure*}

Hence, contour distributions of $c\Omega/V_{tip}$ from the present PIV data for five time-steps spanning the rear fin up-stroke cycle at each of the $\delta$ values are shown in Figure~\ref{fig_Ocontour}.
The vortices associated with the leading and trailing edges of the front fins, are hereafter abbreviated as FLEV and FTEV respectively. Up- and down-strokes are notated using $\uparrow$  and $\downarrow$ respectively. Based on the fin orientations and the present $x$-$z$ view, the signs of their $c\Omega_y/V_{tip}$ are fixed, i.e. FLEV$\uparrow$  and FTEV$\downarrow$ are positive, whereas, FLEV$\downarrow$ and FTEV$\uparrow$  are negative. For convenience, these structures are labeled in the first four time-steps in Figure~\ref{fig_Ocontour} with a legend at the bottom. Depending on whether their sign is positive (counter-clockwise) or negative (clockwise), they are colored red or blue respectively.

For the $\delta$ = -90$^\circ$ case, by the time the rear fin starts its upstroke, the front fin has already reached mid-upstroke and shed the FLEV$\uparrow$ and FTEV$\uparrow$ along with the incumbent FLEV$\downarrow$ that still persists in the vicinity. Throughout its up-stroke, the rear fin is in close proximity to a negatively signed FLEV$\downarrow$+FTEV$\uparrow$, that induce flow downward ($W<$0). 

For the in-phase case, the present generation of structures (FLEV$\downarrow$+FTEV$\uparrow$) is too far from the rear fin LE and the prior generation of flow structures is extremely weak or non-existent (FLEV$\uparrow$). Hence, there is no apparent induced flow from the front fin structures even when the rear fin reaches mid-upstroke.

The $\delta$ = +90$^\circ$ case on the other hand contains a negatively signed FTEV$\uparrow$ for just the beginning of the upstroke. This explains why the rear fin generates the most thrust at this time instant (Figure~\ref{fig_rearfinthrust}b between 860-960 ms). However, soon after, the front fin goes through its mid down-stroke and the dominant structure in the vicinity of the rear fin LE is a strong positively signed FLEV$\uparrow$+FTEV$\downarrow$ structure that induces flow upward ($W>$0), opposite to that of the $\delta$=-90$^\circ$ case. Therefore, the phasing results in the rear fin catching a different front-fin flow structure that substantially affects LE suction for the rear fin and thereby, the thrust generated. 

To further aid visualization of the motion and interaction between flow structures, contour plots of $W/V_{tip}$ from PIV data overlaid with velocity vector field (every third vector shown in each direction for clarity) as well as contour plots of $c\Omega/V_{tip}$ for all ten time-steps are shown in continuous cyclically repeating movies. See Supplemental Material at [URL] for Videos 1-6. The cycle playback time of the movies is increased by 2x for clarity. Velocity maps from PIV data for $\delta$ = -90$^{\circ}$, 0$^{\circ}$ and +90$^{\circ}$ are shown in Videos 1, 2 and 3 respectively, while vorticity maps are shown in Videos 4, 5 and 6 respectively.

\subsection{RLE suction effect}
\begin{figure}[ht]\begin{center}\includegraphics[width=\linewidth]{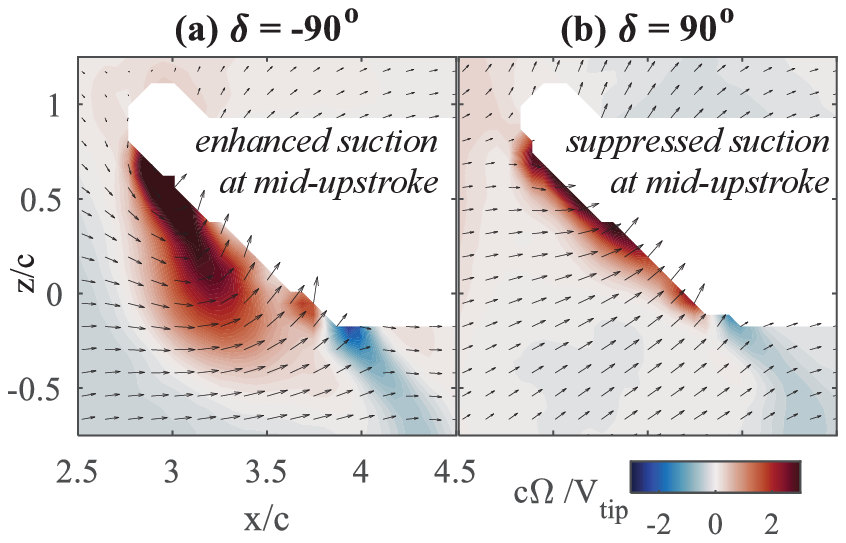}\end{center}
\caption{Contour distributions of $c\Omega/V_{tip}$ from PIV data overlaid with velocity vector field (every third vector shown in each direction for clarity) focusing near the rear fin during its mid-upstroke for (a) $\delta$ = -90$^{\circ}$  and (b) +90$^{\circ}$.}
\label{fig_RLEsuction}\end{figure}

Following Akhtar et al.~\cite{akhtarHydrodynamicsBiologicallyInspired2007}, zoomed-in views of flow near the rear-fin LE when it is at mid-upstroke for the $\delta$ = -90$^\circ$ and +90$^\circ$ cases are shown in Figure~\ref{fig_RLEsuction}a and b respectively. 
Contour distributions of $c\Omega/V_{tip}$ from the PIV data are overlaid with the velocity vector field, with every third vector shown in each direction for clarity. The RLEV$\uparrow$ is distinctly visible from the close-up views. For the $\delta$ = -90$^\circ$ case, the vortex is separated more than $c$ away from the base at $z/c$=0 and there is recovery at $x/c>$3.5. However, for the $\delta$ = +90$^\circ$ case, the RLEV is attached throughout the LE and does not show any signs of separation and subsequent recovery.

An increase in LE separation, such as the one seen in Figure~\ref{fig_RLEsuction}a may be a consequence of either increased inflow/suction or angle of attack. However, from Table~\ref{table002}, we know that $\alpha$ does not change substantially between different $\delta$ cases, indicating that the observed separation is indeed due to increased suction. Therefore, it is evident that the upstream flow conditions significantly enhance and suppress the rear fin LE suction for the $\delta$ = -90$^\circ$ and +90$^\circ$ cases respectively.

Although this effect appears similar to that observed in prior multi-fin studies~\cite{rivalRecoveryEnergyLeading2011,rivalVortexInteractionTandem2011,broeringNumericalInvestigationEnergy2012,boschitschPropulsivePerformanceUnsteady2014}, there is one key difference. The structure that favorably interacts with the rear upstroke is a merged form of FLEV$\downarrow$ and FTEV$\uparrow$, whereas the prior studies attribute this only to the FLEV$\downarrow$. In fact, Rival et al.~\cite{rivalRecoveryEnergyLeading2011,rivalVortexInteractionTandem2011} found the FTEV to be detrimental to the rear thrust, and associated it with the sub-optimal phase offsets. This is due to the fin kinematics and geometry of this study that is relevant for various flapping fin vehicles\cite{lichtDesignProjectedPerformance2004,sitorusDesignImplementationPaired2009,katoBiomechanismsSwimmingFlying2008,palmisanoRoboticPectoralFin2012,gederDevelopmentUnmannedHybrid2017,RazorAUVAC}, as the high pitch angle combined with a spanwise variation of chord results in the FLE and incumbent FTE to be close to each other after every stroke reversal. Therefore, the $\delta$=-90$^\circ$ with the present design leverages two generations of front-fin vortex structures to maximize the thrust at the rear.

\begin{figure}[ht]\begin{center}\includegraphics[width=\linewidth]{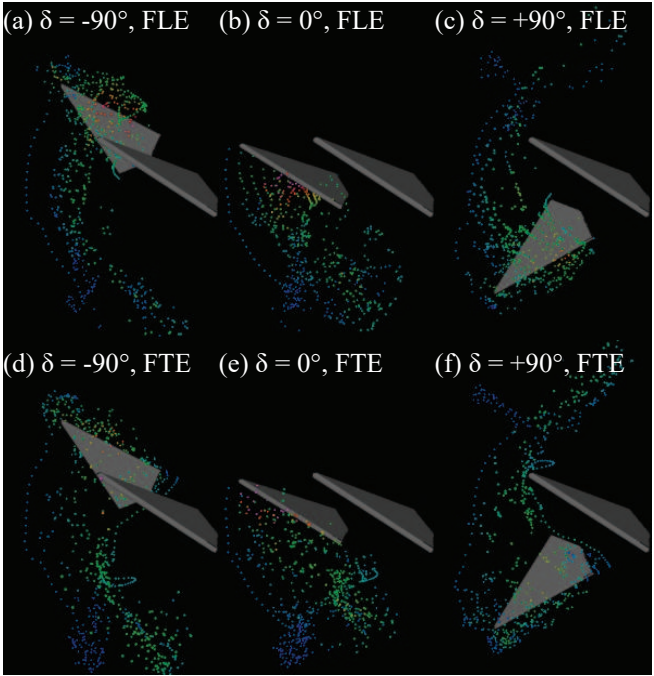}\end{center}
\caption{Snapshot of evolving particle rakes released from the front fin LE and TE acquired from the CFD data  when the rear fin is at mid-upstroke for $\delta$ values (a, d) -90$^{\circ}$ (b, e) 0$^{\circ}$ and (c, f) +90$^{\circ}$ respectively. Particles are color-coded with their velocity magnitude (red: high, blue: low).}
\label{fig_CFDparticlerake}\end{figure}

The results from the 2D-PIV analysis (Figures ~\ref{fig_Wcontour}-~\ref{fig_pressure}) sufficiently explain the effect of phasing on thrust profiles. We now use the CFD results to gain further insight, especially utilizing their fully three-dimensional nature. For instance, Figure~\ref{fig_Ocontour} illustrates the effect of the FLEV and FTEV flow on the rear fin. This is specifically explored further by seeing the evolution of particle rakes released along the FLE and FTE as shown in Figure~\ref{fig_CFDparticlerake} for $\delta$ values of -90$^{\circ}$(a, d),  0$^{\circ}$(b, e) and +90$^{\circ}$(c, f) respectively.
It is evident that only in the $\delta$ = -90$^\circ$ case (Figure~\ref{fig_CFDparticlerake}a, d) the FLE and FTE particle rakes impinge on a substantial portion of the rear fin. In fact, almost the entire span of the rear fin LE has particles in the vicinity. For the in-phase case, there are absolutely no FLE/FTE particles that reach by the time the fin is in its mid-upstroke, consolidating the present claims. For the $\delta$ = +90$^\circ$ case, there are some particles that interact with the RLE tip alone. These particles also have substantially lower velocity magnitude when compared to those interacting with the RLE in the $\delta$ = -90$^\circ$ case.

\subsection{Pressure distributions}
\begin{figure}[ht]
\begin{center}\includegraphics[width=\linewidth]{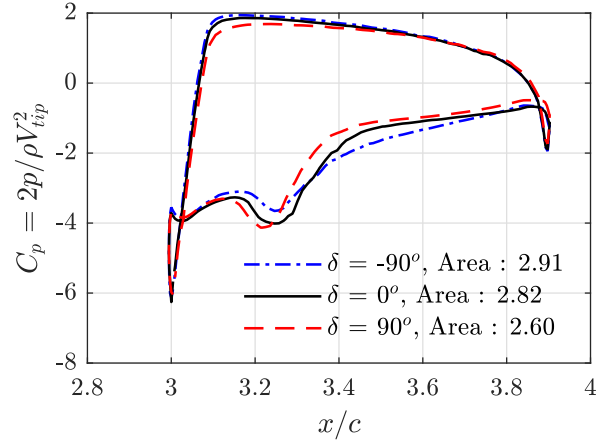}\end{center}
\caption{Profiles of $x/c$ versus $C_p$ along the rear fin acquired from the CFD data when the rear fin is at mid-upstroke for the three $\delta$ values. Area enclosed inside each curve is provided in the legend.}
\label{fig_pressure}
\end{figure}

Following prior studies\cite{akhtarHydrodynamicsBiologicallyInspired2007,rivalRecoveryEnergyLeading2011,rivalVortexInteractionTandem2011}, another comprehensive means of examining the effect of phasing on the rear fin thrust profiles is based on the surface pressure distributions, that are available in the CFD simulations. This encompasses the cumulative three-dimensional effect of all the flow structures on the fins. Hence, profiles of $C_p$ = 2p/$\rho V_{tip}^2$ along the rear fin when it is at the mid-upstroke for all the $\delta$ values are shown in Figure~\ref{fig_pressure} as a function of $x/c$.
The stagnation pressure peak value around $x/c$=3.1 is the highest for the $\delta$=-90$^\circ$ case, followed by the 0$^\circ$ and -90$^\circ$ cases respectively. This is to be expected as the phasing in the -90$^\circ$ case leads to a reduction in $V_{ind}$, thereby reducing the total inflow momentum that is converted to the LE stagnation pressure. The suction side (negative) peak is the closest to the LE and sharpest for the +90$^\circ$ case, located around $x/c$ = 3.22. The flow recovers very quickly around $x/c$ = 3.40. For the in-phase case, the peak shifts to an $x/c$ =3.24 and flow recovers around $x/c$ = 3.47. In contrast, the suction side peak for the -90$^\circ$ case is weaker in amplitude, located around $x/c$ = 3.26 and recovery occurs only at $x/c$ = 3.77, almost near the RTE. This signifies the enhancement in LE suction that stretches the RLEV farther away from the fin in comparison to the +90$^\circ$ case. 

The areas enclosed under the $C_p$ curves also reflect the thrust trends and the profiles clarify exactly where the gains are being made, i.e. at the suction because of delayed recovery as well as the pressure side stagnation point. They represent the sectional thrust per unit span for this particular spanwise location, and are not constant as the fin geometry varies with span. They provide a quantitative sense of the difference between the thrust (for the same spanwise location) at the rear fin for different upstream (front-fin conditions). The total thrust, i.e. when integrated over the entire fin, has been discussed in \ref{thrustprofiles}.

Although a second suction peak such as that observed by Akhtar et al.\cite{akhtarHydrodynamicsBiologicallyInspired2007} is not observed here for the $\delta$=-90$^\circ$ case for this plane cut due to the spanwise structure of the flow, it is the same effect.

\subsection{Wake interaction modes}
A notable contribution by Boschitsch et al.~\cite{boschitschPropulsivePerformanceUnsteady2014} was the identification of the coherent and branched wake interaction modes that exist between 2-D in-line foils. The coherent mode formed a single-core axial jet when the axial velocity was averaged over the entire cycle and was associated with peak thrust and efficiency of the rear fin. The branched mode resulted in a split dual-core jet with the least thrust and efficiency. Kurt and Moored~\cite{kurtFlowInteractionsTwo2018} studied 3-D tandem-fin interactions and found the coherent wake mode to be linked with the highest thrust production, but not the peak propulsive efficiencies. Using 2-D CFD simulations, for closely-spaced tandem fins with small heave amplitudes, Lin et al.~\cite{linPhaseDifferenceEffect2019} defined a `slow' interaction mode that governed the dynamics for tandem fins with larger separations where just a single vortex interaction was observed. The effect was due to the vortex-induced flow causing a lift with the same direction as the flapping motion, similar to the findings reported in Ramananarivo et al.~\cite{ramananarivoFlowInteractionsLead2016}.

The present study has a significantly higher heave amplitude ($h/c$ = 3) making the coherent wake modes~\cite{boschitschPropulsivePerformanceUnsteady2014,kurtFlowInteractionsTwo2018} irrelevant. For a given fin (front or rear), the wakes produced by the up- and down-strokes are separated by a distance greater than 4$c$ (Figure~\ref{fig_Wcontour}) and a significant radially inward migration would be needed to bring the two wakes closer to achieve a so-called `coherent' mode and this doesn't occur for any of the phases. The only phase offset where an inward flux is observed is for the rear fin at $\delta$ = -90$^\circ$ (Figure~\ref{fig_Wcontour}a-e). For the upstream wake to reach $z/c$ = 0 in this case within half a cycle (0.5 s), the vertical velocity would have to be greater than 2$c$/(0.5 s) = 20 cm/s = 0.3$V_{tip}$ which is still 50$\%$ larger than the peak value near $z/c$=2, $x/c$=3 (Figure~\ref{fig_Wcontour}e). Therefore, regardless of phasing, we expect the wake interaction modes to be `branched'. 

\begin{figure}[ht]
\begin{center}\includegraphics[width=\linewidth]{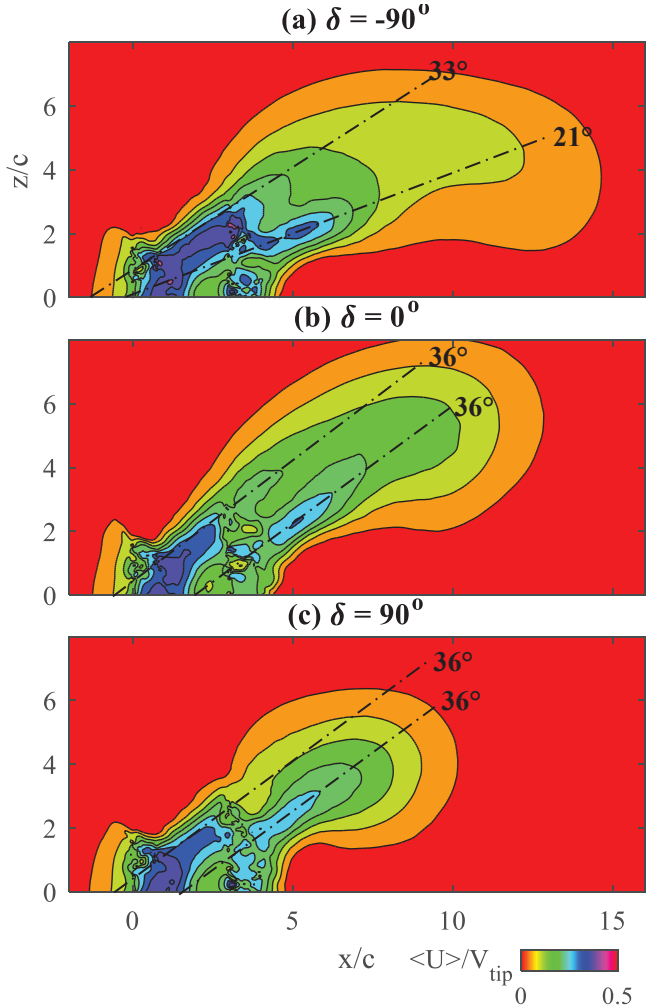}\end{center}
\caption{Contour distributions of $\langle U \rangle/V_{tip}$ covering an extended upper half of the tandem fins extracted from the CFD data for $\delta$ values (a)-90$^{\circ}$ (b) 0$^{\circ}$ and (c) +90$^{\circ}$. }
\label{fig_wakes}
\end{figure}

It is still useful to quantify how `branched' the wakes in the present study actually are and we examine the average axial velocity over the entire stroke cycle, denoted as $\langle U\rangle$ in the upper half region for the three $\delta$ values in Figure~\ref{fig_wakes}a, b and c respectively. The wake inclination angle for the front and rear fin wakes is also labeled for all cases.
As expected, the absence of any axial momentum in the wake close to $z/c$=0 for $x/c>$5 for any of the cases, confirms that the present tandem pectoral fins fall within the `branched' wake mode. Looking first at the extents of the far wakes, there is a clear distinction between each of the three $\delta$ values in Figure~\ref{fig_wakes}. The far wake for the  -90$^{\circ}$ case is the most axially stretched out whereas the +90$^{\circ}$ case is very short-lived. Also interesting to note is that the far wake for the -90$^{\circ}$ case is the least inclined (closest to $z/c$=0) compared to the other two $\delta$ values, indicating that it is the most `coherent' of the three. As described earlier, both these observations are consistent with high thrust production for the $\delta$ = -90$^\circ$ case. The inner region reveals wakes corresponding to the front and rear fins as streaks that are joined together in the -90$^{\circ}$ case and separated in the other two, suggesting that the most favorable interactions are in the former. The front and rear fin wakes are inclined at 36$^\circ$ for the $\delta$ = 0$^\circ$ and +90$^{\circ}$ cases. The fact that they are parallel for these $\delta$ values also suggests the lack of any merging. For the  $\delta$ = -90$^{\circ}$ case, the front fin is inclined slightly lower at 33$^\circ$, and substantially lower for the rear fin at  21$^\circ$. Therefore, a reduction of 15$^\circ$ is achieved by lagging the rear fin by a quarter cycle w.r.to the front fin. 

Due to the zero in-flow condition ($St\rightarrow\infty$) in the present study, radial flow from the front fin has ample space and time to migrate away from the second fin. It is well known that thrust-producing foils at high Strouhal numbers are associated with an inclined wake~\cite{shindeJetMeanderingFoil2013,godoy-dianaModelSymmetryBreaking2009}. For an individual heaving foil, as $St$ increases, the fins go from `drag producing', to `momentum-less' to `thrust-producing'~\cite{godoy-dianaTransitionsWakeFlapping2008,bohlMTVMeasurementsVortical2009,koochesfahaniVorticalPatternsWake1989}. Upon further increase, wake inclination occurs and thrust decreases, motivating studies to employ flexible foils to control the wake asymmetry and optimize thrust performance~\cite{shindeFlexibilityFlappingFoil2014}. Godoy-Diana et al.~\cite{godoy-dianaModelSymmetryBreaking2009} also showed that in these regimes, instead of the reverse von K\'{a}rm\'{a}n or B\'{e}nard-K\'{a}rm\'{a}n vortex streets, vortex pairs (dipoles) are formed and lead to the inclination or branching of the wake, as is evident from Figure~\ref{fig_wakes} for this study as well.

Moreover a high heave amplitude ($h/c$ = 3.15) reduces the possibility of fin interaction even further. Therefore, phasing the rear fin in such cases becomes an effective means of leveraging any momentum on offer from the front. In this context, a reduction in the wake inclination angle of 12$^o$ such as that seen in Figure~\ref{fig_wakes} is substantial. In terms of the in-flow experienced by the rear fin, this is equivalent to going from $St\rightarrow\infty$ to 0.6~\cite{dongWakeTopologyHydrodynamic2006}.

\begin{figure}[ht]
\begin{center}\includegraphics[width=\linewidth]{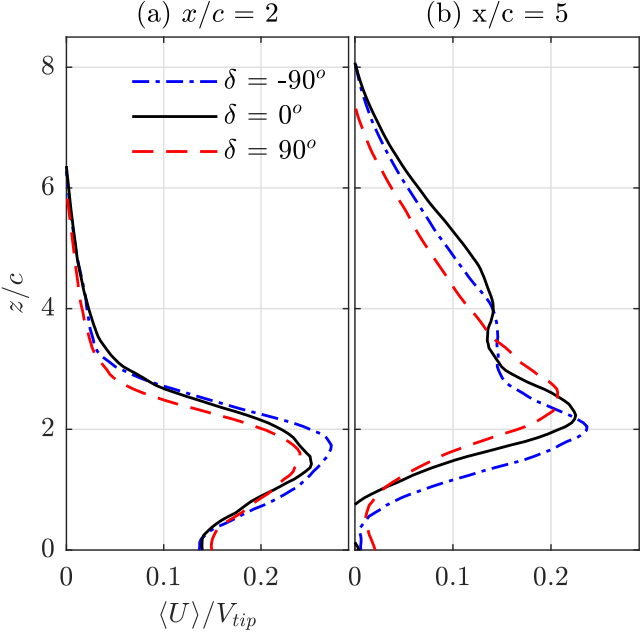}\end{center}
\caption{Profiles of $ \langle U \rangle/V_{tip}$ along axial cuts taken at $x/c$ = 2 and 5 are shown in (a) and (b) respectively.}
\label{fig_wakeprofiles}
\end{figure}

To further evaluate the effect of phasing upstream and downstream of the rear fin, axial cuts of $\langle U\rangle/V_{tip}$ at $x/c$ = 2 and $x/c$ = 5 are shown in Figure~\ref{fig_wakeprofiles}a and b respectively.
The peak values at both axial locations are higher for the -90$^{\circ}$ case, followed by the in-phase case and then the +90$^{\circ}$ case. Although the front fin thrust is unaffected by the phasing of the rear fin (Figure~\ref{fig_thrustresults}), as indicated by Figure~\ref{fig_wakeprofiles}a, there is a small effect of the rear fin phasing that propagates upstream. Figure~\ref{fig_wakeprofiles}b also illustrates the degree of coherence between the three phases, with the -90$^{\circ}$ case being the closest to $z/c$=0 and the +90$^{\circ}$ case the farthest.


\section{Conclusions}
Thrust performance and flow structure characteristics for tandem bio-inspired flapping pectoral fins (trapezoidal with aspect ratio of 3:1) were studied using load cell and 2-D PIV measurements as well as 3-D unsteady viscous CFD simulations. The effects of varying the rear fin stroke phase offset w.r.to the front fin stroke on the thrust and flow characteristics were explored for a fin spacing of 3$c$, heave amplitude of 3.15$c$ and flapping frequency of 1 Hz in an otherwise static water tank ($St\rightarrow\infty$). Three stroke phases were tested, namely when (i) rear fin lagged the front by a quarter cycle ($\delta$ = -90$^\circ$), (ii) rear fin was in phase with the front fin  ($\delta$ = 0$^\circ$) and (iii) rear fin led the front fin by a quarter cycle  ($\delta$ = +90$^\circ$). Key findings of the study are described below:
\begin{itemize}
\item The front fin thrust remained unchanged regardless of the rear-fin phase.
\item The rear fin generated 25$\%$ more thrust on average for the $\delta$ = -90$^\circ$ case, performed the same as the front fin for the $\delta$ = 0$^\circ$ case and produced 8$\%$ lesser thrust for the $\delta$ = +90$^\circ$ case.
\item Phase unwrapping of the rear fin thrust profiles revealed that the maximum thrust gains made by the $\delta$ = -90$^\circ$ case were during mid-stroke with diminishing gains or minor losses around stroke reversal.
\item Stroke-averaged spatial distributions of velocity elucidated the front fin wake and its substantially different interactions with the rear fin for different $\delta$ values. For the $\delta$ = -90$^\circ$ case, the rear fin experienced a downward flow ($W<0$) throughout its upstroke (and vice versa) increasing the effective velocity and suction at the rear fin leading edge (RLE). In contrast, for the $\delta$ = +90$^\circ$ case, the front fin induced an upward flow ($W>0$) for most part of the rear fin upstroke (and vice versa) reducing the effective velocity and suction at the RLE. For the $\delta$ = 0$^\circ$ in-phase case, the core of the front fin wake was farther away from the rear fin, unable to induce any substantial flow on it throughout the cycle. 
\item Estimates of the effective angle of attack, induced velocity and thrust at mid-upstroke of the rear fin showed that there is a definite change in the magnitude of the effective flow but not the flow direction between the three $\delta$ cases, indicating that the thrust gains made by the $\delta$ = -90$^\circ$ are purely due to an increase in suction.
\item Stroke-averaged distributions of spanwise vorticity detail the different interactions that the front fin leading- and trailing-edge vortices (FLEV and FTEV) have with the rear fin for varying $\delta$ values. For the $\delta$ = -90$^\circ$ case, the rear fin is associated with an incumbent clockwise FLEV and the clockwise FTEV from the ongoing front-fin upstroke throughout its own upstroke. These factors substantially enhance the leading-edge separation at the rear fin, which is a well-known marker of the increased suction effect~\cite{akhtarHydrodynamicsBiologicallyInspired2007}. In the $\delta$ = +90$^\circ$ case, a counterclockwise incumbent FLEV and current FTEV suppress suction at the RLE throughout its upstroke.
\item 3-D snapshots of evolving FLE and FTE particle rakes at the same rear fin upstroke phase for the $\delta$ values also show a substantial interaction with the RLE throughout its span for the $\delta$ = -90$^\circ$ case, very little interaction with the RLE for the $\delta$ = +90$^\circ$ case and no interaction for the in-phase case.
\item Chord-wise profiles of the coefficient of pressure along the rear fin in the $x$-$z$ plane at mid-upstroke also confirm the enhancement and suppression of the RLE separation region for the $\delta$ = -90$^\circ$ and +90$^\circ$ cases respectively.
\item Distributions of axial velocity averaged over all cycles show that the present fin geometry and kinematics resemble the `branched' wake interaction modes, with the $\delta$ = -90$^\circ$ wake being the most coherent and closest to the center out of the $\delta$ cases. Both front and rear fin wakes were inclined at 36$^\circ$ in the $\delta$ = +90$^\circ$ and 0$^\circ$ cases. These decreased to 33$^\circ$ and 21$^\circ$ in the $\delta$ = -90$^\circ$ respectively, indicating a marginal upstream effect ($\sim 3^\circ$) and a substantial downstream effect ($\sim 15^\circ$) of phasing on the wake inclination.
\end{itemize}
In conclusion, this study provided detailed perspectives on how rear fin phasing modulates thrust and wake characteristics for tandem flapping pectoral fins. In hovering applications where this is of direct relevance, $\delta$=-90$^\circ$ would correspond to the maximum thrust and the most radially coherent wake `signature', whereas at $\delta$=+90$^\circ$ the axial extent of the wake can be substantially reduced by sacrificing thrust performance. Guided by this study, phasing can be used as a powerful vehicle control tool to achieve desired thrust and wake characteristics on-the-go without making larger-scale system modifications such as fin geometry~\cite{ramamurtiPropulsionCharacteristicsFlapping2019}, stiffness~\cite{shindeFlexibilityFlappingFoil2014}, etc. In addition, the methods and analyses in this study can be extended to study flow mechanisms for various fin geometries and configurations to aid in design and control of a wider range of flapping fin vehicles.

\begin{acknowledgments}
This research was performed while Kaushik Sampath held an NRC Research Associateship award at the U.S. Naval Research Laboratory. The authors wish to acknowledge Dr. Ruben Hortensius (TSI Inc.) for invaluable technical support with the PIV hardware, Dr. Ellen R Goldman (Code 6920, NRL) for providing lab space and equipment to prepare fluorescent seeding particles, Drs. Charles A. Rohde and David C. Calvo (Code 7165, NRL) for graciously lending PIV equipment and Robert S. Leary (Excet Inc./Code 7165, NRL) for assistance with fabrication of mounting equipment. This work was supported by ONR through an NRL 6.2 base program. Computational time for this work was supported in part by a grant of HPC time from the DoD HPC center at NRL.
\end{acknowledgments}

\appendix
\section{PIV data processing, uncertainty and fin vibrations}
\label{AppA}
\begin{figure*}[ht]\begin{center}\includegraphics[width=\linewidth]{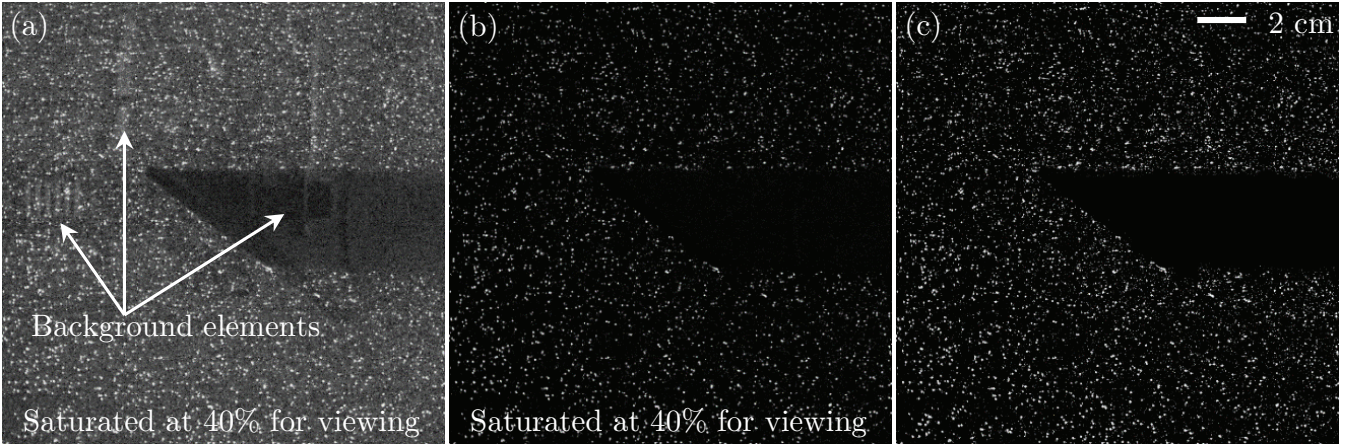}\end{center}
\caption{Steps involved in the image enhancement procedure. (a) Raw image, (b) after stroke-averaged background subtraction, (c) after modified histogram equalization (MHE) enhancement and range-filter based masking. Note that images shown in (a) and (b) are saturated at 40$\%$ for viewing.}
\label{fig_PIVimages}\end{figure*}

Although the use of fluorescent particles eliminates a significant fraction of the background, the large field of view capture lends itself to issues such as a spatially varying image intensity distribution, as well as faint secondary reflections of the mounting hardware from the light scattered by the fluorescent particles. Furthermore, as the fins are optically opaque, in cases when the front fin intersects the $x$-$z$ plane containing the PIV laser sheet, it leaves a shadow downstream, covering portions of the field of view around the rear fin. Although optical refractive index matching between the fins and the working fluid can be used to overcome this limitation~\cite{budwigRefractiveIndexMatching1994,sampathPhaseLockedPIV2015,sorannaEffectsInletGuide2010}, we opt out of it for the present study. Consequently, accurate masking of the shadow regions becomes important. A sample raw image is shown in Figure~\ref{fig_PIVimages}a. 
As a first step, the stroke-cycle averaged background is calculated for the two exposures and subtracted from the raw image, eliminating most of the support structure from the image; see Figure~\ref{fig_PIVimages}b. Modified histogram equalization (MHE) enhancement~\cite{rothFiveTechniquesIncreasing2001} is performed to extend the intensity range in every 128$\times$128 sub-region to its full dynamic range. Evidently, this step removes spatial non-uniformity and enhances the signal-to-noise ratio of the particle intensity distributions. Next, a range filter is applied to group the enhanced particles together, forming a smooth dynamic logical complement of the region that needs to be masked. The resulting masked image, such as the sample shown in Figure~\ref{fig_PIVimages}c is used for the subsequent steps. 

PIV analysis is performed using gauCorr~\cite{rothFiveTechniquesIncreasing2001}, a cross-correlation based code that has been adopted successfully to analyze PIV data in applications ranging from turbo-machines~\cite{sampathPhaseLockedPIV2015,sorannaEffectsInletGuide2010}, oceanic boundary layers~\cite{nayakWaveCurrentInteraction2015} to cardiovascular flows ~\cite{sampathOptimizedTimeResolvedEcho2018}. A cross-correlation window of 32$\times$32 pixels with a 50$\%$ overlap between windows is selected, resulting in a spacing of 3.1 mm between neighboring velocity vectors. Outliers from the resulting vector fields are removed following the universal outlier detection scheme~\cite{westerweelUniversalOutlierDetection2005}. Based on a sub-pixel estimate of displacement-correlation peaks~\cite{adrianParticleImageVelocimetry2011} an uncertainty in the instantaneous velocity of 0.2 pixel/frame = 1.91 cm/s may be obtained, as long as there are five to eight particles per interrogation window. When averaged over 1300 – 2000 strokes, the uncertainty drops to 0.05 cm/s. At this scale however, unsteadiness of flow structures or fin vibrations (discussion follows) may have a larger contribution. 

\begin{figure}[ht]\begin{center}\includegraphics[width=\linewidth]{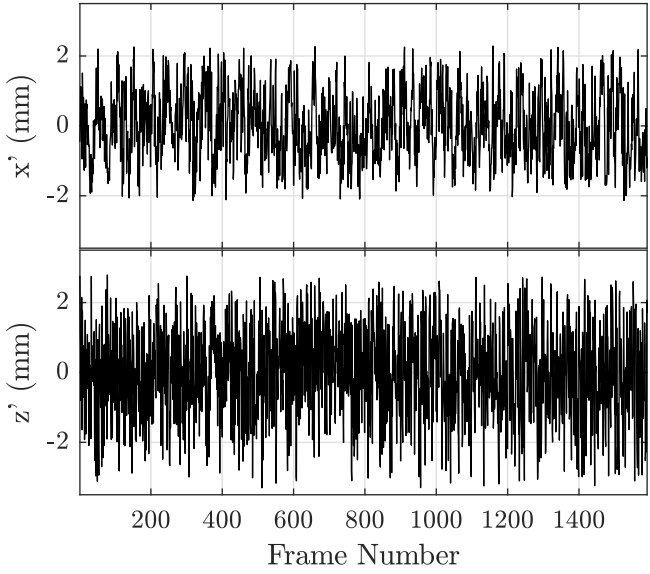}\end{center}\caption{Mean subtracted axial (x’) and vertical (z’) front fin leading edge tip positions for the $\delta$ = -90$^o$, t = 60 ms case.}\label{fig_LEtipvibration}\end{figure}

Although mounted rigidly to an aluminum-extrusion skeleton, the fins are prone to vibrations. Figure~\ref{fig_LEtipvibration} shows the variation in mean subtracted axial and vertical positions of the front fin LE tip through a sample dataset ($\delta$ = -90$^o$, t = 60 ms) that, as discussed earlier, is recorded at exactly the same $\phi_{front}$ each time. Evidently, there are vibrations up to a maximum of $\pm$2 mm, with those in the vertical direction being larger than the axial direction. This is expected as the tip is farther from the stroke axis (= 3.62$c$) compared to its distance from the pitch axis (= 0.25$c$). The vibration amplitudes are roughly one grid spacing (= 3.1 mm) in the PIV velocity distributions. The same trend (not shown) also holds for the rear fin. To test the effect of the blade vibrations on the stroke cycle-averaged velocity distributions, the displacement profile shown in Figure~\ref{fig_LEtipvibration} can be used to translate the image coordinates in ($x$, $z$) and impose a fixed location for the front fin LE to re-calculate the stroke cycle-averaged velocity distributions. Comparisons of the velocity magnitudes between the uncompensated and compensated cases (not shown) indicate that there is no substantial effect of the vibration on the averaged distributions, with the essential features (shape and strength) being the same.

The most important factor is the unsteady nature of the flow structures that makes the scale of the present vibrations inconsequential. Furthermore, the front and rear fins vibrate independently and randomly about their respective locations, with negligible coupling. Therefore, two `different' compensations based on the front and rear fin positions need to be performed, where combining the two datasets becomes non-trivial. Hence, for the present study, we opt to neglect the effect of fin vibrations and do not compensate for it. A static ablated mask (expanded by one spatial grid point) is applied on the entire dataset to ensure there are sufficient number of data-points available when ensemble averaging along the fin edges.

\begin{figure}[h]\begin{center}\includegraphics[width=\linewidth]{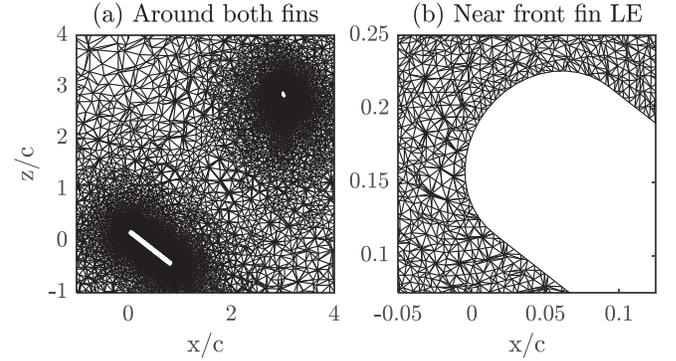}\end{center}
\caption{Snapshots of the CFD mesh for $\delta$ = -90$^o$.}
\label{fig_CFDresolution}\end{figure}

\begin{figure}[ht]\begin{center}\includegraphics[width=\linewidth]{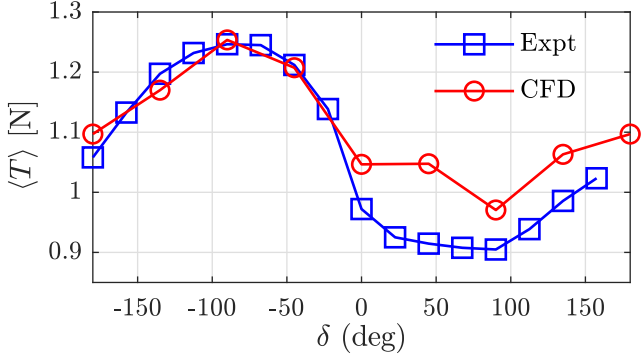}\end{center}
\caption{Average rear fin thrust profiles, $\langle T\rangle$ [N] for different $\delta$ values for experimentally measured thrust and CFD results at $U_\infty$ = 0.}
\label{fig_thrust_delta}\end{figure}

\begin{figure}[ht]\begin{center}\includegraphics[width=\linewidth]{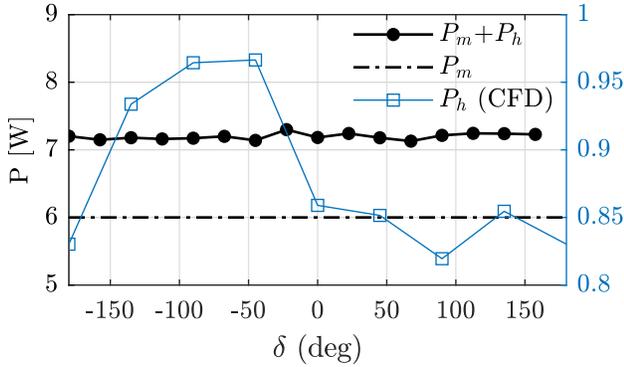}\end{center}
\caption{Rear fin power over different $\delta$ values, measured in water ($P$=$P_m$+$P_h$), in air ($P$=$P_m$) and using CFD ($P$=$P_h$, right axis). Please note the different in scales between the left and right axes.}
\label{fig_power}
\end{figure}

\begin{figure}[ht]\begin{center}\includegraphics[width=\linewidth]{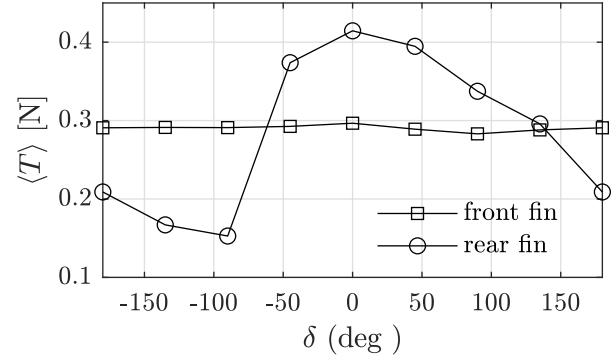}\end{center}
\caption{Average thrust profiles, $\langle T\rangle$ [N] for different $\delta$ values from CFD simulations performed at $U_\infty$ = 0.5 m/s for the front and rear fins.}
\label{fig_Uinf}\end{figure}

\section{Comparison of CFD results with thrust and PIV measurements}
\label{appendix_CFDValidation}

\begin{figure*}[ht]\begin{center}\includegraphics[width=\linewidth]{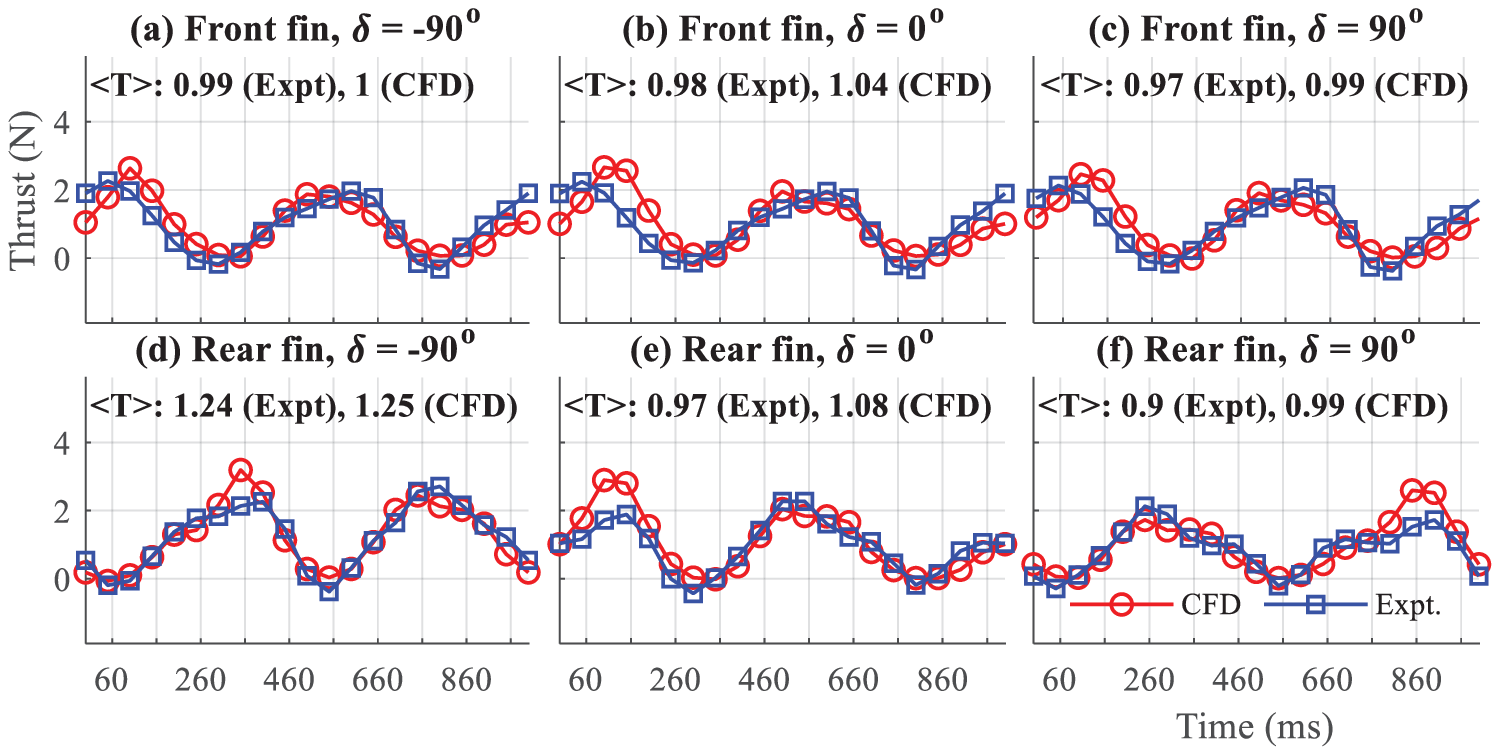}\end{center}
\caption{Stroke-cycle averaged profiles of experimentally measured and frequency filtered axial thrust (N) compared with CFD results for the front (a-c) and rear (d-e) fins for (a, d)  $\delta$ = -90$^{\circ}$, (b, e) $\delta$ = 0$^{\circ}$ and (c, f) $\delta$ = +90$^{\circ}$. }
\label{fig_thrustCFDvalidation}\end{figure*}

The measured axial thrust profiles from Figure~\ref{fig_thrustresults} are compared with the results obtained from the CFD data and shown in Figure~\ref{fig_thrustCFDvalidation}.
The profiles for the front and rear fin are shown in the top (a-c) and bottom rows (d-f) respectively, while the left (a, d), middle (b, e) and right (c, f) columns correspond to the values of $\delta$ = -90$^{\circ}$, 0$^{\circ}$ and +90$^{\circ}$. The trends, as well as the magnitudes of the CFD thrust profiles are very close to those of the experiments for the front and rear fins. The only noteworthy deviations are near mid-upstroke, where the CFD over-predicts the thrust peaks marginally. This is attributed to the lack of smoothness of the fin kinematics that is fed to the simulations directly from the experimentally measured data.

\begin{figure*}[ht]\begin{center}\includegraphics[width=\linewidth]{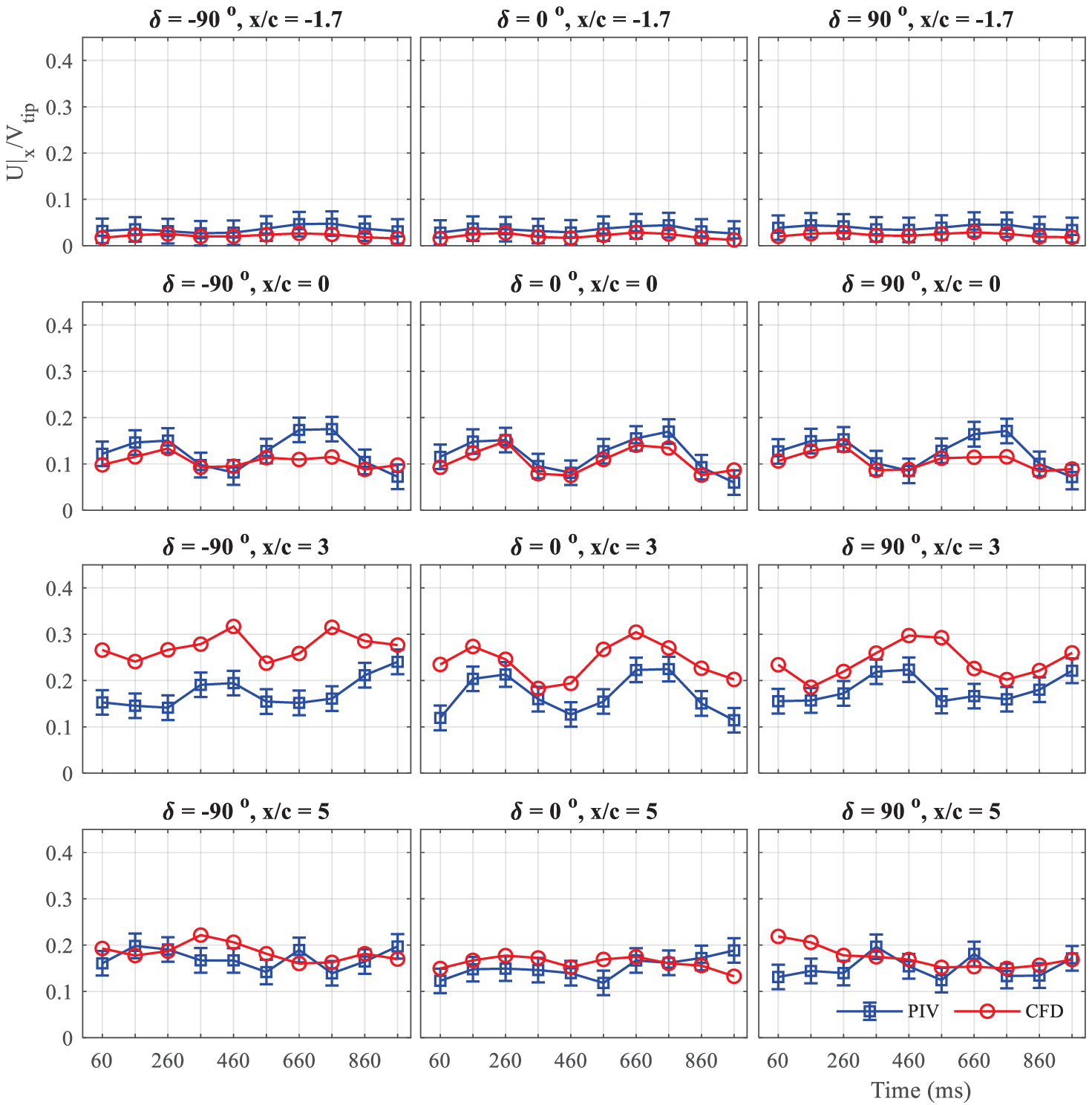}\end{center}
\caption{Evolution of $U|_x/V_{tip}$ averaged over all $z$ at $x/c$ = -1.7, 0, 3 and 5, for $\delta$ = -90$^{\circ}$, 0$^{\circ}$ and +90$^{\circ}$ cases over ten equally spaced time-steps (t = 60, 160…960 ms) spanning one stroke cycle for the PIV ($\square$) and CFD ($\circ$) data. Error bars denote uncertainty in instantaneous PIV data.}
\label{fig_CFDPIVUx}\end{figure*}

\begin{figure*}[ht]\begin{center}\includegraphics[width=\linewidth]{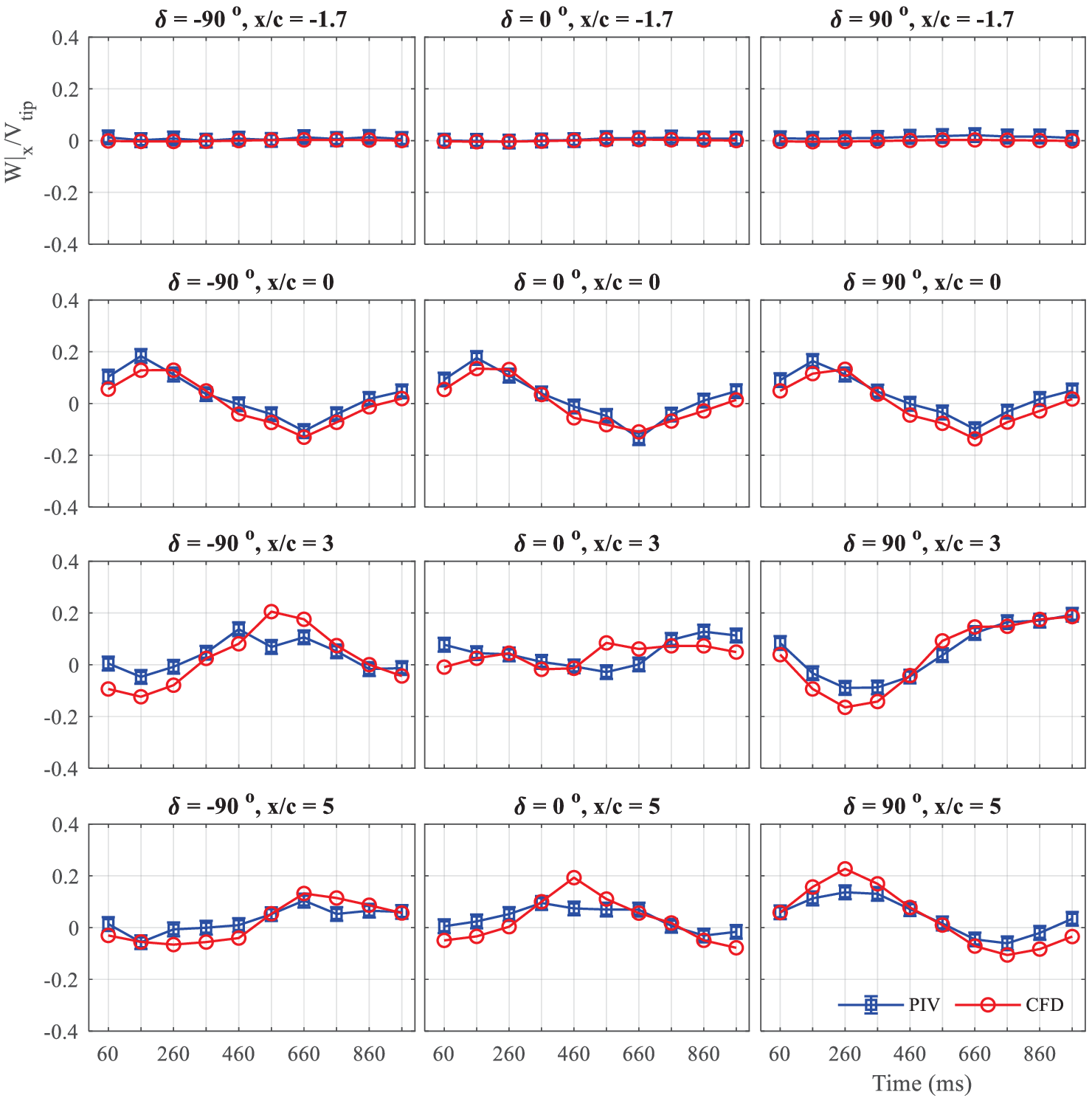}\end{center}
\caption{Evolution of $W|_x/V_{tip}$ averaged over all $z$ at $x/c$ = -1.7, 0, 3 and 5, for $\delta$ = -90$^{\circ}$, 0$^{\circ}$ and +90$^{\circ}$ cases over ten equally spaced time-steps (t = 60, 160, …960 ms) spanning one stroke cycle for the PIV ($\square$) and CFD ($\circ$) data. Error bars denote uncertainty in instantaneous PIV data.}
\label{fig_CFDPIVWx}\end{figure*}

To allow a direct quantitative comparison between the CFD and PIV data for all time-steps profiles of $U/V_{tip}$ and $W/V_{tip}$ are averaged along $z$ at $x/c$ = -1.7, 0, 3 and 5 for the ten equally spaced time-steps (t = 60, 160…960 ms) and shown in Figure~\ref{fig_CFDPIVUx} and Figure~\ref{fig_CFDPIVWx} respectively. 
They are denoted as $U|_x/V_{tip}$ and $W|_x/V_{tip}$ respectively. Results from both PIV ($\square$) and CFD ($\circ$) are included for all the cases. When averaging along $z$, missing (NaN) points (masked/shadowed regions) are not included. The PIV data has more missing points than CFD, making the two profiles inherently different. $\delta$ = -90$^{\circ}$, 0$^{\circ}$ and +90$^{\circ}$ cases are shown in the left, middle and right columns respectively, while the five axial cuts are distributed row-wise. Axial cuts are chosen strategically to highlight trends upstream ($x/c$ = -1.7), at the front fin LE ($x/c$ = 0), rear fin LE ($x/c$ = 3) and downstream ($x/c$ = 5). Error bars for the PIV are chosen based on the instantaneous uncertainty (1.91 cm/s) derived in~\ref{AppA}, although this is definitely an overestimate.

Evidently, despite the contribution of masked points to the PIV data, the overall trends between the CFD and PIV data agree well for both the axial and vertical velocity profiles. The local maxima and minima for each case agree well as well as the amplitudes are in close agreement, except at near the rear fin ($x/c$=3), where the PIV results are substantially affected by masking. Also, the relative trends between different $\delta$ cases are in good agreement, e.g. the profiles of $W|_x/V_{tip}$ at $x/c$=3 and 5.

\clearpage
\bibliography{References}
\end{document}